\documentclass[aps,pra,amsfonts,superscriptaddress,twocolumn]{revtex4} 
\usepackage{epsfig}
\usepackage{bbm} 
\usepackage{multirow}
\usepackage{lscape}
\usepackage{color}
\usepackage{rotating}
\usepackage{hyperref}
\newcommand{\eq}{\begin{eqnarray}} 
\newcommand{\en}{\end{eqnarray}}
\def\bra#1{\mathinner{\langle{#1}|}}
\def\ket#1{\mathinner{|{#1}\rangle}}
\newcommand{\braket}[2]{\langle #1|#2\rangle}
\newcommand{\nicefrac}[2]{{#1}/{#2}}

\usepackage{amsmath} 
 \usepackage{bbm}
\usepackage{placeins}

\begin{document}

\title{Limits to multipartite entanglement generation with bosons and fermions}
\author{Malte C. Tichy}
\affiliation{Physikalisches Institut, Albert--Ludwigs--Universit\"at Freiburg, Hermann--Herder--Strasse~3, D--79104 Freiburg, Germany}
\affiliation{Department of Physics and Astronomy, University of Aarhus, DK--8000 Aarhus C, Denmark}
\author{Florian Mintert}
\affiliation{Freiburg Institute for Advanced Studies, Albert-Ludwigs-Universit\"at, Albertstrasse 19, 79104 Freiburg, Germany}
\affiliation{Physikalisches Institut, Albert--Ludwigs--Universit\"at Freiburg, Hermann--Herder--Strasse~3, D--79104 Freiburg, Germany}
\author{Andreas Buchleitner}  
\affiliation{Physikalisches Institut, Albert--Ludwigs--Universit\"at Freiburg, Hermann--Herder--Strasse~3, D--79104 Freiburg, Germany}

\date{\today}
\begin{abstract}
Many-photon interference in linear-optics setups can be exploited to generate and detect multipartite entanglement. Without recurring to any inter-particle interaction, many entangled states have been created experimentally, and a panoply of theoretical schemes for the generation of various classes of entangled states is available. Here, we present a unifying framework that accommodates the present experiments and theoretical protocols for the creation of multiparticle entanglement via interference. A general representation of the states that can be created is provided for bosons and fermions, for any particle number, and for any dimensionality of the entangled degree of freedom. Using this framework, we derive an upper bound on the generalized Schmidt number of the states that can be generated, and we establish bounds on the dimensionality of the manifold of these states. We show that -- at the expense of a smaller success probability -- more states can be created with bosons than with fermions, and give an intuitive interpretation of the state representation and of the established bounds in terms of superimposed many-particle paths.
\end{abstract}

\maketitle

\section{Introduction}
The preparation of entangled states of photons \cite{Pan:2011fk} has been the subject of a great collective effort, which was driven by fundamental questions such as the violation of local realism \cite{Bell:1964pt,Aspect:1982ly,Aspect:1981zr} as well as by the perspective of technological applications,  e.g.,~quantum cryptography \cite{Ekert:1991kx} and quantum computing \cite{Walther:2005qf}. The control over  spontaneous parametric down-convertion (SPDC) started this successful journey in 1988, when a two-photon entangled state  was created \cite{PhysRevLett.61.2921}. Step-by-step-wise, the number of photons was further increased, such that entanglement in three-photon \cite{Bouwmeester:1999ys}, four-photon \cite{Eibl:2003ly}, five-photon \cite{Zhao:2004dz}, six-photon \cite{Lu:2007ve,Radmark:2009ij,Prevedel:2009ec,Wieczorek:2009ff}, and, eventually, eight-photon states \cite{Yao:2011uq,Huang:2011fk} was demonstrated. While many experiments can generate only one particular state, some setups  can tune the final state through several classes of multipartite entanglement \cite{RevModPhys.81.865,Wieczorek:2009fe}. 

The creation of an entangled state of photons relies on the independent propagation of the particles in a linear-optics setup and on the subsequent selection of successful final states with one particle in each spatial output mode, i.e.~on \emph{post-selection}. No direct interaction between the photons takes place, in contrast to other possible carriers of entanglement such as  trapped ions \cite{Blatt:2008cl}. It is, in contrast, the \emph{indistinguishability} of the photons which provides the seed for quantum correlations. Therefore, the physical mechanism -- the coherent superposition of many-particle paths -- is rather independent of the specific implementation and of the particle species -- bosonic or fermionic. For example, also the indistinguishability of cold bosonic or fermionic atoms can be used for interaction-free entanglement generation \cite{Popescu:2007kx}.

Besides polarization, also other degrees of freedom can  carry photonic entanglement, such as the time-bin \cite{PhysRevLett.82.2594,Marcikic:2004qr,Halder:2007th}, the path \cite{PhysRevLett.94.220501,PhysRevLett.61.2921}, or the orbital angular momentum \cite{Vaziri:2002nx,Inoue:2009dq,Mair:2001zr,PhysRevA.83.012306}. These degrees of freedom  are not restricted to two dimensions, but allow the manipulation of high-dimensional \emph{qudit}-entanglement. In particular, qutrit-entanglement was realized with energy-time entangled photons \cite{Thew:2004ly}, and access to a  20-dimensional time-bin-like photonic degree of freedom was demonstrated \cite{PhysRevA.69.050304}. 

Manifold further strategies for the creation of entangled states were proposed \cite{PhysRevA.80.022308,PhysRevLett.99.193602,Maser:2009pi,Schilling:2009fk,PhysRevLett.102.053601,Maser:2010fu,Lim:2005bf}, which promise to carry on the experimental achievements  \cite{Pan:2011fk,Wieczorek:2009ff,Wieczorek:2008uq,Wieczorek:2009fe,Prevedel:2009ec}; but although all schemes share the same physical ingredients, no common framework permits a direct comparison. With the  increasing complexity of many-photon  setups, a general framework is highly desirable to achieve  high success rates, a low number of components, and a comparison of competing approaches to  entanglement generation \cite{Maser:2010fu,Lim:2005bf}. A  unified description is therefore our first desideratum in this article.

Conceptually, we also lack an understanding for the intrinsic limitations of setups that rely on many-particle interference.  At first sight, it seems that a  large set of entangled states of photons of nearly arbitrary particle number and dimension can be created in the laboratory. It was, however, also  recognized that a toolbox composed of linear optics and post-selection alone is by far not universal, and several no-go theorems were formulated: With linear optics, a Bell measurement that discriminates all four Bell-states is impossible  \cite{Lutkenhaus:1999bh}, as well as  deterministic teleportation  \cite{Vaidman:1999qf}. On the other hand, the manipulation and characterization of qudits is severely constrained: The Schmidt-rank of any measurement operator on two qudits is bounded by $N$, when $N-2$ auxiliary qudits are available \cite{Calsamiglia:2002dq}.  Furthermore, it is not known whether bosons and fermions differ in their capability for entanglement generation, e.g. whether all states that can be created with photons can also be generated using fermions in some analogous setup. The computational complexities of many-fermion and many-boson scattering processes differ substantially \cite{Aaronson:2010fk}, which suggests that bosons are superior to fermions also in the present context. While  a treatment of fermions is not immediately necessary for the characterization of photonic entanglement, the understanding of possible differences between fermions and bosons may also help to better appreciate the full potential of photonic entanglement generation.

While the characterization of multipartite entangled states is rapidly progressing \cite{PhysRevA.83.022328,Bastin:2009ye,PhysRevLett.104.020504,Bin-Liu1:2011fk,salwey,PhysRevLett.106.180502}, it becomes desirable to understand the general physical principles that limit their creation,  and a common framework  will certainly  improve our understanding significantly. Such framework should, both, describe the physical processes that lead to a certain final state, and offer a representation of this state that is compatible with the tools of quantum information theory, like entanglement measures. In other words, such framework should bridge the gap between the \emph{physical} many-particle scattering process and the \emph{abstract} quantum state of  quantum information theory.

Here, we set the stage for the  investigation of the aforementioned problems.  We propose a treatment of entanglement-generating setups which accommodates virtually all possible schemes that use identical particles.  The created state can then be understood  as a coherent superposition of many-particle paths \cite{1367-2630-14-9-093015,tichydiss}, which can be traced back to the state representation. In particular, we find a  combinatorial bound for the generalized Schmidt number \cite{Eisert:2001uq}, which generalizes a  theorem for two particles \cite{Calsamiglia:2002dq}. This bound applies equally for bosons and fermions, for which, however, the dimensionality of the set of states that can be created turns out to be different:  more states can be generated with bosons than with fermions.

We first establish the  experimental capabilities that are available, in Section \ref{formalism}. On that basis, we formulate the requirements that a versatile model should fulfill, and develop  an adequate  framework. An explicit representation of the achievable  states is  derived in Section \ref{repre}, with  which we formulate  a  bound on the Schmidt number of the generated states.  We then show in Section \ref{Dimsetsse} that  the number of dimensions of the manifold of states that can be created with bosons and fermions establishes another limitation. Concluding discussions are given in Section \ref{concluout}. For illustration, and in order to familiarize the reader with  our framework, we schematically reproduce several state-of-the-art experiments and theoretical schemes in Appendix \ref{classification}. The derivation of the dimensional bounds is presented in Appendix \ref{appdimbounds}.

\section{Unified framework for entanglement generation} \label{formalism}

\subsection{Physical situation and versatility requirements}
The examples for entanglement generation that were mentioned in the introduction are all based on the same principle mode of operation: Several photons prepared in a given initial state propagate through a linear-optics setup. Post-selection is performed on events at the output modes, and  only events with exactly one particle in each spatial mode are taken into account.  Given such a post-selected, successful event, the internal states of the particles have then the desired entanglement properties.  The presence of a photon in a mode is usually ensured by detection of the particle, which destroys the state; but also non-destructive, heralded preparation is possible with additional auxiliary photons  \cite{Pittman}. Besides entanglement generation via propagation and post-selection, also the projection onto entangled states and the detection of entangled states is possible  \cite{Wieczorek:2009fe,Halevy:2011cr}, which permits, e.g., to transfer photonic entanglement to massive particles \cite{Monroe}. 

Given the rather broad variety of existing schemes and successful experiments, our dedicated  attempt to formalize the state-of-the-art under one housetop should fulfill the following requirements:
\begin{itemize}
\item Any number of particles $N$ is allowed. 
\item Entanglement is established between arbitrary $d$-dimensional degrees of freedom. 
\item The coherent manipulation of the entangled degrees of freedom is possible without restriction.
\item The initial state may or may not be entangled. 
\item Bosons as well as fermions may carry the entanglement. 
\item The creation of entanglement as well as the projection onto entangled states are contained in the description. 
\end{itemize}

In contrast to the polarization of photons, which can be manipulated essentially without restrictions, it is not immediate that all coherent single-particle operations or state-dependent beamsplitters can be implemented for any other degree of freedom \cite{Zou:2005hc}. Depending on the chosen physical carrier of entanglement --  e.g.~polarization, time-bin, path, or orbital angular momentum --, it may be possible that, e.g.,~operations that discriminate the internal degrees of freedom are  difficult to implement \cite{Zou:2005hc,FRANKEARNOLD208}. It is therefore important to pay particular attention to the modeling of different types of setups with distinct capabilities, as we will do in Section \ref{treatment}.

\subsection{Many-particle states and their interpretation}

A central building block of our present approach is the description of the physical many-particle scattering process and its relationship to the  entangled states it generates. We emphasize the conceptual difference between the abstract, quantum-information interpretation of a state (the ``$N$-qudit-state''), and its \emph{physical realization}, which we will need to work with in parallel. 

For the understanding of an experiment, it is useful to describe \emph{physical states} of photons or other identical particles in  second quantization; we will denote these states by  $\ket{\Psi^{2q}}$. In this representation, we can formulate the physical many-particle scattering process, whereas the definition of an entanglement content requires the identification of a subsystem structure which is most adequately done in first quantization \cite{Tichy:2011fk}. Within a clear tensor structure that is imposed by physical observables, the usual means to characterize entanglement, such as monotones and measures \cite{Ghne:2009ys}, can then be applied on the first-quantized state $\ket{\Phi^{1q}}$.

Since the dynamics is best described in second quantization, whereas first quantization is more appropriate for the description of entanglement properties, we will specify a map, $\hat \Omega_a$, between both representations such that $ \ket{\Psi^{2q}}=\hat \Omega_a \ket{\Phi^{1q}}.$  We emphasize that the conceptual problem of the assignment of entanglement to a state of identical particles \cite{Ghirardi:2003uq,Ghirardi:2004fk,PROP:PROP201200079} is circumvented by the present treatment, since states of identical particles $\ket{\Psi^{2q}}$ are identified with states of distinguishable subsystem components $\ket{\Phi^{1q}}$. Before we can define this map, let us briefly review the essential properties of quantum states. 

\subsubsection{State representation} \label{staterepsec}
An $N$-qudit-state,
\eq \ket{\Phi^{1q}(c)}& =&\sum_{j_1=1}^d \dots \sum_{j_N=1}^d c_{j_1,\dots, j_N} \ket{j_1, \dots , j_N}  ,  \label{initstates} \en
is fully specified by the $d^N$ components of the coefficient tensor \eq c  \in \mathbbm{C}^d \otimes \mathbbm{C}^d \otimes \dots \otimes \mathbbm{C}^d. \label{Ctod} \en 

The subsystem structure in (\ref{initstates}) and (\ref{Ctod}) reflects the paradigmatic situation of $N$ parties that  each control a $d$-dimensional subspace. Therefore, the usual notions of separability and entanglement can be applied on  $ \ket{\Phi^{1q}(c)}$ \cite{RevModPhys.81.865,Nielsen:2000fk,Werner:1989ve}. For instance, the state $ \ket{\Phi^{1q}(c)}$ is fully separable (or $N$-separable) if $c_{j_1, \dots, j_N}$ can be written as a product, 
\eq 
c_{j_1,\dots , j_N} = c^{(1)}_{j_1} \cdot c^{(2)}_{j_2} \cdot \dots \cdot c^{(N)}_{j_N} . \label{separablet}
\en
The $N$-qudit-state $\ket{\Phi^{1q}(c)}$ is realized in the experiment as a physical state $\ket{\Psi^{2q}(c)}$ of $N$ bosons or fermions that each carry a $d$-dimensional degree of freedom. Since the carriers of entanglement are indistinguishable particles, each subsystem is of the same dimension $d$.  This degree of freedom will typically be an internal property of the particle, such as polarization or orbital angular momentum. In the following, we will always refer to entanglement between these internal degrees of freedom when speaking of entanglement between particles. Additionally, the particles are distinguished by some external, discriminating degree of freedom - typically the spatial mode. Our description thus requires $N$ spatial modes (as many as particles), such that the minimal size of the single-particle Hilbert space for a description in second quantization has dimension $n=d\cdot N$, since each particle can live in any of the $N$ modes, due to indistinguishability. In order to incorporate setups that treat different internal degrees of freedom in a distinct way (the internal state of an incoming particle may change during the scattering process), we choose a notation in which the internal (entangled) and external (discriminating) degrees of freedom are not explicitly distinguished, which will considerably simplify the subsequent discussion. For this purpose, the first $d$ physical modes (of a total of $n$ modes) are interpreted as the $d$ internal states of a particle that is prepared in the first spatial mode, the second $d$ modes represent the $d$ internal states of a particle in the second spatial mode, and so on. In other words, we consider the entanglement between the particles found in groups of $d$ modes. 

The creation operator $\hat a_{k,l}^\dagger$ creates a particle in the $k$th spatial mode and in the $l$th internal state, and we identify 
\eq 
\hat a^{\dagger}_{k,l}  \equiv \hat a_{d (k-1)+l}^\dagger . \label{identific}
\en 

With this convention, a second-quantized state possesses an interpretation in terms of $N$ (first quantized) particles that are controlled by $N$ parties -- the usual quantum information paradigm of an $N$-qudit-state -- if and only if there is exactly one particle per group of $d$ modes. We call such physical states that can be interpreted as $N$-qudit-state \emph{post-selected states}, since these are precisely the states that are obtained by post-selection. Any $N$-qudit-state in first quantization possesses a representation in second quantization, but not vice-versa.

The second-quantized version of the initial $N$-qudit-state $\ket{\Phi^{1q}(c)}$ in (\ref{initstates}) is thus necessarily a post-selected state, 
 \eq 
\ket{\Psi^{2q}(c)}=\sum_{j_1=1}^d \dots \sum_{j_N=1}^d c_{j_1,\dots, j_N} \prod_{k=1}^N \hat a_{d\cdot (k-1)+j_k}^\dagger \ket{\text{vac}} , \label{secnd111} \en
where $\ket{\text{vac}}$ denotes the vacuum. The occupation of the $n=d\cdot N$ physical modes is characterized by a \emph{mode occupation list}, 
\eq \vec r= (r_1, r_2, \dots, r_{n} ) \label{molist}
,\en which describes a state with $r_j$ particles in the $j$th mode. 
 Equivalently, we can specify the modes in which the $N$ particles are located, i.e.~define a \emph{mode assignment list}. For each term in the state (\ref{secnd111}), the latter reads \eq 
\vec d(\vec r) = (j_1, d+j_2, 2d+j_3,\dots , (N-1)d +j_N ) . \label{masslist}
\en
We illustrate the mode occupation and mode assignment lists of several distinct physical states and the corresponding $N$-qudit-states in Table \ref{scheme2p}.

\begin{table}[h]

\centering
\begin{tabular}{rrrrr} 
 $d$ &~ $N$ &~  occupation~$\vec r$ & ~assignment~$ \vec d(\vec r)$ &  $\ket{\Phi^{1q}}$  \\  \hline \hline
 3 & 2 & $(\overbrace{1,0,0},\overbrace{0,1,0}) $& (1,5) & $\ket{1,2}$ \\
 2 & 3 & $(\overbrace{1,0},\overbrace{1,0},\overbrace{0,1}) $& (1,3,6) & $\ket{1,1,2}$ \\
 2 & 4 & $~(\overbrace{1,0},\overbrace{0,1},\overbrace{0,1},\overbrace{1,0}) $& (1,4,6,7) &~ $\ket{1,2,2,1}$ \\
 2 & 3 & $(\overbrace{1,0},\overbrace{1,1},\overbrace{0,0}) $& (1,3,4) & no \\ 
 3 & 2 & $(\overbrace{1,0,1},\overbrace{0,0,0}) $& (1,3) & no \\ 
 2 & 2 & $(\overbrace{2,0},\overbrace{0,0})$ & (1,1) & no \\  
       \end{tabular} 
       \caption{ Physical Fock states and interpretation as $N$-qudit-states $\ket{\Phi^{1q}}$. The Fock states are defined by the mode occupation list $\vec r$ [Eq.~(\ref{molist})], or, equivalently, by the mode assignment list $\vec d(\vec r)$ [Eq.~(\ref{masslist})]. If exactly one particle is present in each group of $d$ modes (which are emphasized by the brackets), there is a corresponding first-quantized $N$-qudit-state $\ket{\Phi^{1q}}$. The last three states  correspond to unsuccessful scattering events and cannot be used for quantum-information purposes within our scheme.  } 
\label{scheme2p}
\end{table}

\subsubsection{Map}
As anticipated above, we define an explicit map $\hat \Omega_{a}$ between the  first-quantized $N$-qudit-state and the  second-quantized representation, 
\eq 
\hat \Omega_a = \sum_{j_1=1}^{d} \dots  \sum_{j_N=1}^{d}  \left[ \prod_{k=1}^N \hat a^\dagger_{d(k-1)+j_k } \right]
 \ket{\text{vac}} \bra{j_1, \dots , j_N }^{1q} , \label{omegadd} 
 \label{omegaddm1} \en 
 such that 
 \eq 
\ket{\Psi^{2q}(c) } &=& \hat  \Omega_a \ket{\Phi^{1q}(c)} , \nonumber \\  \ket{\Phi^{1q}(c) } &=& \hat  \Omega_a^\dagger \ket{\Psi^{2q}(c)} =\hat  \Omega_a^{-1} \ket{\Psi^{2q}(c)}   \label{secondqstate} .\en
The map $\hat \Omega_a$ is defined for a given particle-number $N$, and invertible only for the subset of post-selected many-particle states. The post-selection procedure is described by the projector $\hat P_{1,a}$ on the subspace of states with exactly one particle in each group of $d$ modes,  
\eq \hat P_{1,a} = \sum_{j_1, \dots, j_N=1}^{d} \left( \prod_{k=N}^1 \hat a^\dagger_{  d(k-1)+j_k} \right)  \left( \prod_{k=1}^N \hat a^{\phantom \dagger}_{ d(k-1)+j_k} \right)  \label{interpretableproject}  ,\en
where the inverted order of the indices for the creation operators avoids sign changes due to the exchange of fermions. For example, the first three states given in Table \ref{scheme2p} are eigenstates of $\hat P_{1,a}$ and possess an interpretation as $N$-qudit-state, whereas the last three do not and lie in the kernel of $\hat P_{1,a}$.

\subsection{Many-particle evolution}
The manipulation of entanglement by the thus defined setup relies on the scattering of non-interacting particles that carry the possibly entangled internal degree of freedom. In other words, the time-evolution of a  physical initial state $\ket{\Psi^{2q}(c)}=\hat \Omega_a \ket{\Phi^{1q}(c)}$ 
is governed by a single-particle scattering matrix $W$ that connects $n$ input modes $\hat a_j^\dagger$ to $n$ output modes $\hat b^\dagger_k$. The state of each particle prepared in an input mode thus evolves into a superposition of amplitudes localized at different output modes: 
\eq \label{wscatteringeq} \hat a_j^\dagger \rightarrow \sum_{k=1}^n W_{j,k} \hat b_k^\dagger .  \en 
The physical scattering process is governed by the Schr\"odinger equation, it is thus unitary. The scattering matrix $W$, however, is non-unitary, in general, since particle loss may occur, i.e.~a particle that is prepared in an input mode has a finite amplitude to not reach any output mode. The matrix $W$ is, in general, embedded in a unitary matrix of dimension $(2n-1)\times(2n-1)$ \cite{He:2007tg,Bernstein:2009uq}, such that 
\eq
U=\left(\begin{array}{cc} 
 W & X \\
Y & Z 
\end{array} \right) , \label{embedma}
\en
where $X, Y, Z$ are matrices of dimension $n \times (n-1)$, $(n-1) \times n$ and $(n-1) \times (n-1)$, respectively, i.e.~at least $n-1$ auxiliary modes are needed to account for particle loss. The non-vanishing matrix elements of $X$ describe the loss of particles, i.e.~they quantify the transition amplitudes that describe the processes in which an injected particle is not collected by any of the first $n$ modes. For example, particles may be lost in free-space propagation schemes \cite{Maser:2009pi,PhysRevLett.99.193602,Ma:2012kx}. 
The embedding matrix $U$ can be inverted, and the inverse scattering process on the $n$ modes of interest  (with the input and output modes exchanged) is thus described  by $W^\dagger$, and 
\eq  \hat b_j^\dagger \rightarrow \sum_{k=1}^n W^\dagger_{j,k} \hat a_k^\dagger = \sum_{k=1}^n W^\star_{k,j} \hat a_k^\dagger .  \label{reversescat}   \en
That is to say, even though $W$ does formally not necessarily possess an inverse, the scattering process with exchanged input- and output modes is always described by $W^\dagger$.

\begin{figure}[h]
\center
\includegraphics[width=\linewidth,angle=0]{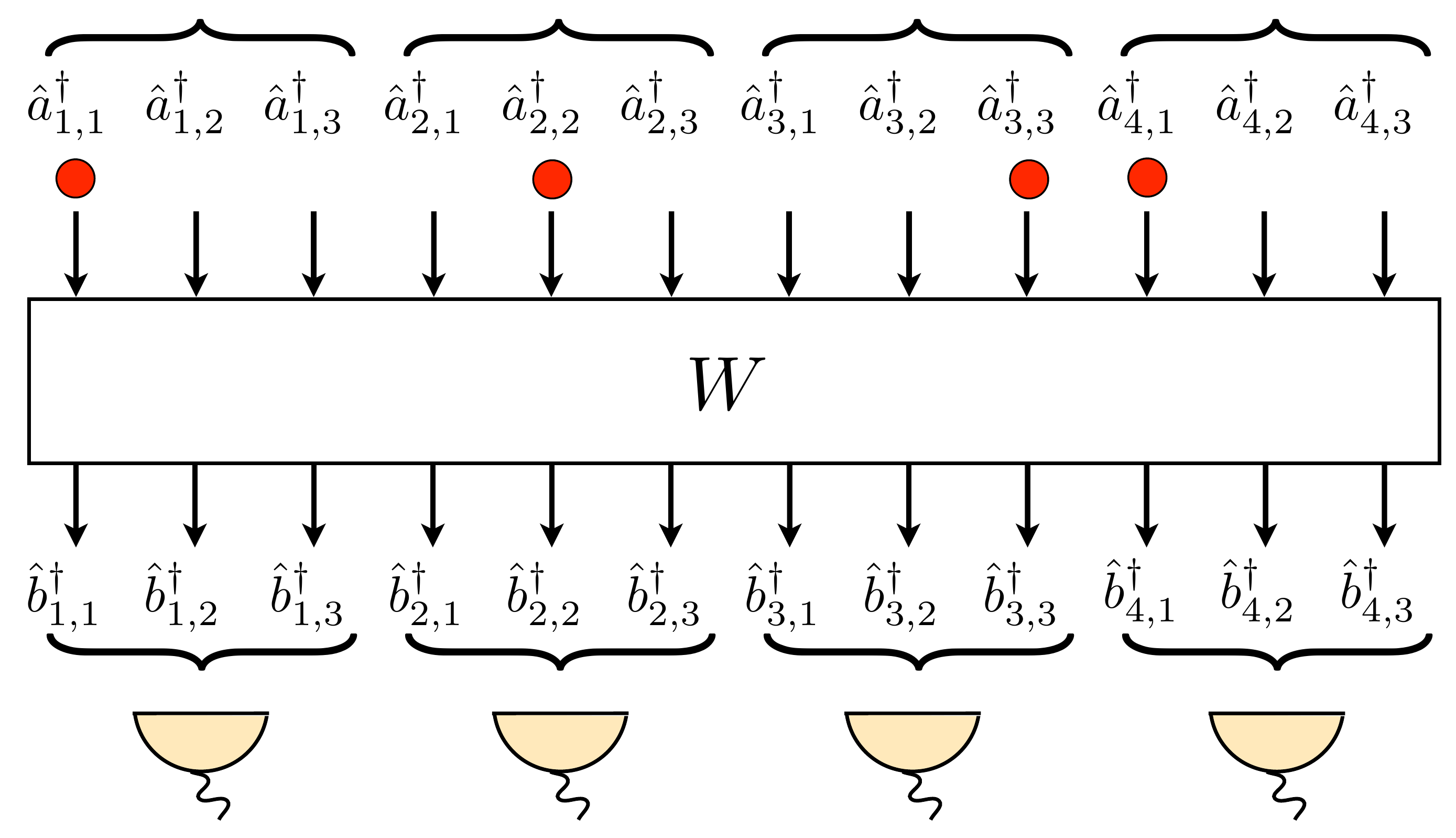} 
\caption{(color online)   Scattering setup for $N=4$ qutrit-like ($d=3$) particles.  The input state corresponds to $\vec r=(1,0,0,0,1,0,0,0,1,1,0,0)$, $\vec d(\vec r)=(1,5,9,10)$, and $\ket{\Phi^{1q}(c)}=\ket{1,2,3,1}$. } \label{schemepicture}
\end{figure}

The single-particle evolution governed by the matrix $W$ applies for every single particle. The many-particle state $\ket{\Psi^{2q}(c)}$ thus evolves according to the time-evolution operator $\hat M(W)$, which maps a Fock state of particles that were prepared in the input modes, i.e.~a state given in terms of input-mode creation operators $\hat a_{k}^\dagger$, to a state in terms of output creation operators $\hat b_{j}^\dagger$. It thus describes the effect of the single-particle evolution $W$ on the many-particle state. Inserting (\ref{wscatteringeq}) into (\ref{secnd111}), we find the effect of $\hat M(W)$, 
\eq 
\hat M(W) \ket{\Psi^{2q}(c)} \hspace{5.6cm} \label{maeleMW} \\ \nonumber   = \sum_{j_1=1}^d \dots \sum_{j_N=1}^d c_{j_1,\dots, j_N} \prod_{k=1}^N  \left(\sum_{l=1}^n W_{d\cdot (k-1)+j_k,l} \hat b^\dagger_l   \right)\ket{\text{vac}} . 
\en 
For each summand of the outer multi-sum (which runs over $j_1, \dots j_N$), we find 
\eq  \label{bigsum}
 \prod_{k=1}^N  \left(\sum_{l=1}^n W_{d\cdot (k-1)+j_k,l} \hat b^\dagger_l   \right)=\hspace{3.23cm} \\  \nonumber   
=\sum_{\vec r} \left[  \sum_{\sigma \in S_{\vec d(\vec r)}} \text{sgn}_{B/F}(\sigma)  \prod_{m=1}^N W_{d(m-1)+j_m,\sigma(m)}  \right] \prod_{l=1}^n \left( \hat b^\dagger_{l} \right)^{r_l}
\en
where the sum over $\vec r$ runs over all possibilities, given $W$, to distribute the particles among the final modes, $\vec d(\vec r)$ is the mode assignment list, and $\sigma$ runs over all permutations of this list. The fermionic/bosonic nature of the particles is taken into account by $ \text{sgn}_{\text{B/F}}(\sigma) $, which is unity for bosons (B) or the signature of the permutation $\sigma$ for fermions (F). The time-evolved state $\hat M(W) \ket{\Psi^{2q}(c)}$ is thus a coherent superposition of all the possibilities to distribute the particles among the output modes. Similarly to single-particle paths that contribute to an event that exhibits interference, \emph{many-particle} paths are superposed coherently \cite{1367-2630-14-9-093015}. We illustrate the process with a scattering setup for $N=4$ particles that each can populate $d=3$ internal states, in Fig.~\ref{schemepicture}. 

\subsubsection{Properties of the final state}
After scattering through the setup, the final, second-quantized state (\ref{maeleMW}) consists of a coherent superposition of all particle arrangements $\vec r$, with corresponding amplitude given by (\ref{bigsum}). In general, not all components can be interpreted as $N$-qudit-states, since  double occupation of some mode or some group of $d$ modes may occur.  We thus need to project the state onto the subspace of post-selected states via the application of $\hat P_{1,b}$, and subsequently apply the inverse mapping $\hat \Omega_b^{-1}$, where $\hat \Omega_b$ is defined in full analogy to $\hat  \Omega_a$ in (\ref{omegadd}), but with interchanged input- and output-creation operators $\hat a_k^\dagger$ and $\hat b_k^\dagger$. We eventually obtain the final $N$-qudit-state \eq \ket{\Phi^{1q}(\tilde g)}= \hat \Omega_b^{-1} \hat P_{1,b} \hat M(W) \hat \Omega_a \ket{\Phi^{1q}(c)}, \en with a norm that equals the success probability of the procedure (in the following, unnormalized tensors will be denoted as such by a tilde). The normalized state $\ket{\Phi^{1q}(g)}$ describes an $N$-partite state of qudits, and it is fully defined by its coefficient tensor  $g_{j_1\dots j_N}$ in the representation analogous to (\ref{initstates}), which 
allows us to characterize its entanglement properties with the tools of quantum information theory \cite{RevModPhys.81.865,Nielsen:2000fk}.

\begin{figure}[h]
\center
\includegraphics[width=\linewidth,angle=0]{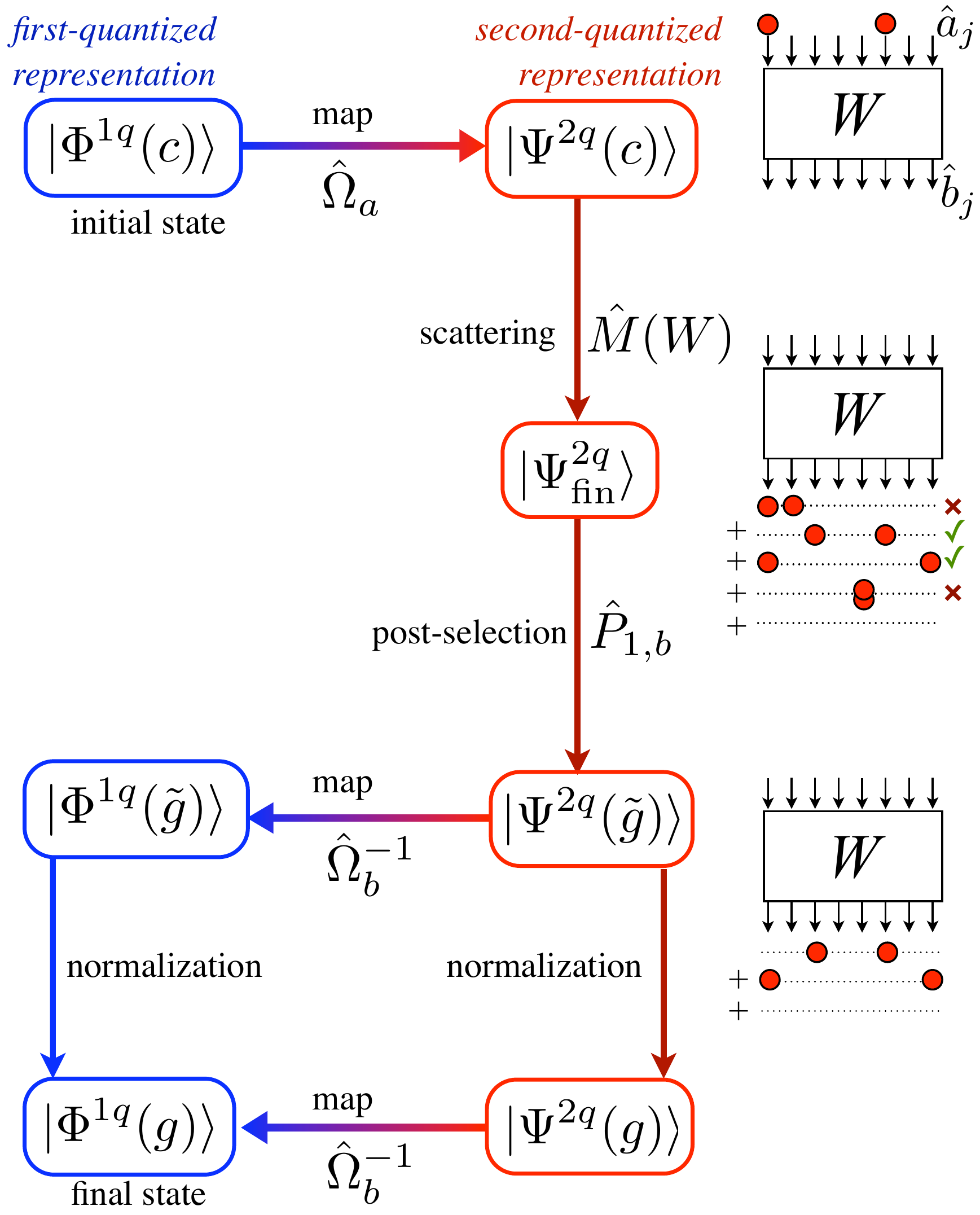} 
\caption{(color online)   Entanglement manipulation via many-particle scattering. Given the $N$-qudit-state 
$\ket{\Phi^{1q}(c)}$, Eq.~(\ref{initstates}), the physical second-quantized state is given by Eq.~(\ref{secondqstate}), $\ket{\Psi^{2q}(c)}=\hat \Omega_a \ket{\Phi^{1q}(c)}$, with the map operator $\hat \Omega_a$ given by Eq.~(\ref{omegadd}). The scattering of the particles through the setup is described by $\hat M(W)$,  Eq.~(\ref{maeleMW}), the projection on the subspace of post-selected states is mediated by $\hat P_{1,b}$, Eq.~(\ref{interpretableproject}). The mapping to a first-quantized state occurs via $\hat \Omega_b^{-1}$, Eq.~(\ref{omegaddm1}).   The sketch on the right-hand side shows the physical situation for $d=4$ and $N=2$. The initial state (upper sketch)  contains only one component, it is injected into the scattering setup. The final state contains many coherently superposed components. Some of them (marked by red crosses) are not interpretable as $N$-qudit states. The projector $\hat P_{1,b}$ describes the postselection mechanism, i.e., it eliminates the unwanted components and leaves only the postselected components (green crotchets). } \label{SchemeSecondFirst}
\end{figure}

\subsection{Entanglement creation and detection} \label{CreationDetection}
The above-defined mapping between first- and second-quantized representation and the general procedure for the manipulation of entanglement via many-particle scattering are illustrated in Fig.~\ref{SchemeSecondFirst}. As we show now, the process can be used as a protocol for entanglement generation, detection and projection:
\begin{description}
\item[Entanglement generation.] Given an initial separable state $\ket{\Phi^{1q}(c)}$, a target entangled state 
\eq \ket{\Phi^{1q}(\tilde g)}=\hat \Omega_b^{-1} \hat P_{1,b} \hat M(W) \hat  \Omega_a \ket{\Phi^{1q}(c)} \en is generated by scattering and post-selection. The success probability of the entanglement generation process is given by $ \braket{\Phi^{1q}(\tilde g)}{\Phi^{1q}(\tilde g)}= \left|\hat P_{1,b} \hat M(W) \hat  \Omega_a \ket{\Phi^{1q}(c)} \right|^2 .$ For example, Refs.~\cite{PhysRevLett.61.2921,Lim:2005bf,Maser:2010fu} rely on this procedure. 
\item[Entanglement detection.] Given an entangled initial state $\ket{\Phi^{1q}(c)}$, the scattering process yields a superposition that contains one or several separable (known) \emph{signal states} $\ket{\Phi^{1q}(s^{(k)})}$, i.e.~$\ket{\Phi^{1q}(\tilde g)}=\Omega_b^{-1} \hat P_{1,b} \hat M(W) \Omega_a \ket{\Phi^{1q}(c)}$ with $\braket{\Phi^{1q}(s^{(k)})}{\Phi^{1q}(\tilde g)}\neq 0$. All states that are orthogonal to the initial state do not lead to any signal state, such that the detection of signal states is a signature for the initial entangled state. 
\item[Entanglement projection.] If we start with a superposition of a desired target entangled state $\ket{\Phi^{1q}(g)}$ -- as for the entanglement detection scheme -- with other undesired components, the detection of a separable signal state $\ket{\Phi^{1q}(s^{(k)})}$ projects the initial superposition onto the target state. This procedure is used, for example, in entanglement swapping protocols \cite{Halder:2007th}, or to project atoms into entangled states by measuring the photons emitted by the atoms \cite{PhysRevLett.99.193602}. 
\end{description}

The three tasks pose very distinct experimental requirements in general, and they are therefore typically not realized in the same experimental setup. Still, they are fully equivalent in theory: For any number of particles $N$ and any dimensionality of the entanglement carrier $d$, every created state can also be detected, and every state that can be detected can also be created, in principle. 

More formally, a target entangled state $\ket{\Phi^{1q}(g)}$ that is created via the scattering of an initial, separable state $\ket{\Phi^{1q}(c)}$ on a setup $W$ can be detected with the very same separable state as the signal state, via the scattering of the target entangled state on $W^{\dagger}$, i.e.~using the output modes of $W$ as input. 

In order to show this, we need to ensure that $\ket{\Psi^{2q}(c)}$ qualifies as a signal state. We thus require (i) a finite detection probability,
\eq \bra{\Psi^{2q}(c)} \hat M(W^\dagger)  \ket{\Psi^{2q}(g)} \neq 0 . \label{creatdd} \en 
Furthermore, (ii) for all states characterized by $g^\bot$ that are orthogonal to the target state $g$, the signal state is never produced, i.e. 
\eq \forall g^\bot, \text{ with } \braket{\Phi^{1q}(g^\bot) }{\Phi^{1q}(g)}=0: \nonumber \\ \bra{\Psi^{2q}(c)} \hat M(W^{\dagger}) \ket{\Psi^{2q}(g^\bot)} = 0 . \label{projectdd} \en
Eqs.~(\ref{creatdd}) and (\ref{projectdd}) are actually satisfied as a direct consequence of 
 $\hat M(W^\dagger)=\hat M^\dagger(W)$, i.e.~by the fact that the inverse scattering process is described by $W^\dagger$ (see (\ref{embedma}) and (\ref{reversescat})). 

In full analogy, any state that can be detected can also be created. In order to facilitate our notation in the following, we will always refer to the \emph{entanglement generation} of a desired \emph{target entangled state} -- our results can then be applied analogously for entanglement detection and projection.

\subsection{Treatment of internal degrees of freedom} \label{treatment}
The principle mode of operation being established, it remains to specify the properties of the scattering matrix $W$ for specific physical systems. 

The formalism can accommodate setups in which particles are simply redistributed among output modes, without any manipulation of their internal degrees of freedom. We will refer to such setups as \emph{non-polarizing setups} (which does not necessarily imply the restriction to polarization as the internal degree of freedom). Setups that are sensitive to the internal degree of freedom, i.e.~\emph{polarizing setups}, are also thinkable (such as polarizing beam-splitters, which redirect photons  conditioned on their polarization, but which do not change the polarization of the incoming particles). We call a setup that  also manipulates the polarization, e.g.~by using wave plates, \emph{polarization-manipulating setup}. For these three different types of setups, the scattering matrix $W$ will assume a distinct structure, and distinct target states are accessible. 

\subsubsection{Non-polarizing setups} \label{Nonpolarizingsec}
When photons propagate in free space or scatter off a non-polarizing multiport beam-splitter that treats all internal degrees of freedom equally, we have a propagation of the form 
\eq \hat a_{j,l}^\dagger \rightarrow \sum_{k=1}^N V_{j,k }\hat b_{k,l}^\dagger , \en 
for particles that are prepared in a spatial mode $j$ and in the internal degree of freedom $l$, where we distinguished internal and external degrees of freedom for the moment (See Eq.~(\ref{identific})). Each degree of freedom is treated independently and equally, which formally translates into a scattering matrix $W$ that contains one elementary scattering matrix $V$ of dimension $N \times N$, 
\eq
W&=&V \otimes \mathbbm{1}_d .
 \label{nonpolmatrix}
 \en
 For example, the matrix that describes a non-polarizing beam-splitter, 
\eq W&=&\frac{1}{\sqrt 2} \left(\begin{array}{cccc} 
1 & 0 & 1 & 0 \\
0 & 1 & 0 & 1  \\
1 & 0 & -1 & 0 \\
0 & 1 & 0 & -1   \end{array} \right) = 
\frac{1}{\sqrt 2} \left(
\begin{array}{cc}
1 & 1 \\
1 & -1  \end{array}
 \right) \otimes \mathbbm{1}_2 \label{nonpolbs}  ,\en possesses precisely the structure (\ref{nonpolmatrix}).  Any scheme that does not manipulate the internal degree of freedom of the scattered particles, such as the ones discussed in \cite{Maser:2010fu,Schilling:2009fk,Maser:2009pi,PhysRevLett.99.193602,Lim:2005bf}, can be accomodated by a matrix of the form (\ref{nonpolmatrix}).

The entanglement content of the final $N$-qudit-state depends on the prepared initial internal states of the particles: The multiset of internal states of the particles is not altered, i.e.~an initial state $\ket{\Phi^{1q}}=\ket{\epsilon_1, \epsilon_2 , \dots \epsilon_N}$ will lead to a final state that can be expressed by superpositions of terms which consist of permutations of the $\ket{\epsilon_k}$ on the output modes. If all $\ket{\epsilon_k}$ are equal, no entanglement can be created. 

\subsubsection{Polarizing setups} \label{polasetups}
The situation is similar for \emph{polarizing setups}, where we allow different dynamics for each internal state, but no coupling between them, e.g.,~a horizontally polarized photon may be directed to another output mode than a vertically polarized photon, but the polarization of incoming photons is never changed. This model is realistic for degrees of freedom that can be discriminated, but which cannot be  manipulated at will, i.e.~the device equivalent  of a wave plate is not available. The scattering matrix is still a direct sum of scattering matrices, which, however, do not need to be equal,
\eq W &=& V^{(1)} \otimes \text{diag}[(1,0\dots , 0)] \nonumber \\ 
&& + V^{(2)} \otimes \text{diag}[(0,1,0\dots , 0)] \nonumber \\
&&  + \dots \nonumber \\ && +  V^{(d)} \otimes \text{diag}[(0,\dots 0,1)] 
\label{polasetupBBB} , \en
where the $V^{(k)}$ are $N \times N$ matrices and $\text{diag}[\vec v]$ is the matrix that contains $\vec v$ on the diagonal, whereas all other elements vanish. 

For example, a polarizing beam-splitter does not modify polarization, but its action on the modes depends on the polarization state. The matrix 
\eq
W_{\text{PBS}} &=&
\frac 1 {\sqrt 2} \left(\begin{tabular}{cccc} 
1 & 0 & 1 & 0 \\
0 & 0 & 0 & $\sqrt{2}$ \\
1 & 0 & -1 & 0 \\
0 & $\sqrt{2}$ & 0 & 0 
\end{tabular}
 \right)   \\
&  =  & \frac 1 {\sqrt 2} \left(\begin{tabular}{cc} 
1 & 1 \\
1 & -1 
\end{tabular} \right) \otimes \text{diag}[(1,0)] \nonumber \\
&& + \left(\begin{tabular}{cc} 
0 & 1 \\
1 & 0 
\end{tabular} \right) \otimes \text{diag}[(0,1)] \nonumber ,
\en
describes a polarizing beam-splitter which reflects and transmits vertically polarized photons with the same probability and always transmits horizontally polarized photons. 

\subsubsection{Polarization-manipulating setups}
Finally, a setup which does not only treat different internal degrees of freedom in a distinct way, but also allows their manipulation is described by an unrestricted scattering matrix $W$. This is experimentally more demanding in general, but routinely done for photon polarization, with the help of phase shifters, polarization rotators (e.g.~implemented by the combination of quarter- and half-wave-plates) and polarizing beam-splitters. 

\subsection{Effective degrees of freedom and independence of initial state} \label{initialstateindepen}

\subsubsection{Multiple mode occupation}
At first sight, the formalism seems to restrict severely the types of setups with which entanglement is produced: Often, the \emph{initial} state in an experiment does not possess an interpretation as $N$-qudit-state in the quantum information abstraction, such that no $\ket{\Phi^{1q}}$ can be given. For example, a many-particle state with all particles in the same spatial mode is scattered on several beam-splitters in \cite{Wieczorek:2009ff}, which permits the creation of a permutation-symmetric entangled state. 

Such initial multiple population is, however, unproblematic: If two particles are prepared in the same mode, we can equally well describe them by a setup in which they formally occupy two \emph{distinct}  modes, but for which the two rows of the matrix representation that describes the evolution of the particles are identical. We can thus absorb such initial preparation of particles into the matrix $W$. 

\subsubsection{Local unitary operations} \label{localunitaries}
All local operations that act independently on each particle in the input or in the output modes, i.e.~local unitary operations \cite{Nielsen:1999uq,Vidal:2000kx}, can also be absorbed by the scattering matrix: 
\eq W= \left( L^{(i)}_1 \oplus   \dots  \oplus L^{(i)}_N  \right) W_0 \left( L^{(f)}_1 \oplus \dots  \oplus L^{(f)}_N \right) , \label{genscattwithlocal} \en 
where $L^{(i)}_k$ and $L^{(f)}_l$ are $d\times d$-matrices that describe local unitary transformations on the $k$th spatial input- and on the $l$th spatial output mode, respectively. We can therefore restrict our analysis to a unique initial state $\ket{\Phi^{1q}}=\ket{1,1,1,\dots,1}$, in which all particles are prepared in the same internal state and in distinct spatial modes. All relevant parameters can be varied by the entries of the scattering matrix $W$. 
 
\section{State representation and combinatorial bounds} \label{repre}
Having established the physical process for entanglement generation as well as the possible structures of the scattering matrix $W$, we now proceed to the explicit representation of the states that can be generated, given an initially separable state. The form for polarization-manipulating setups will be the most general one, while we find a simpler representation for non-polarizing setups. In all cases, elementary algebraic manipulations allow us to obtain a form from which the maximal Schmidt rank, i.e.~the maximal number of separable components of a state, can be read off. 

\subsection{Non-polarizing setups and permutation representation}
For a non-polarizing setup, we are given a scattering matrix of the form (\ref{nonpolmatrix}), with an $N\times N$-matrix $V$ as non-trivial ingredient. Any separable initial state can be written as
\eq 
\ket{\Phi^{1q}(c)}=\ket{\epsilon_1, \dots , \epsilon_N} ,
\en
where $\ket{\epsilon_k}$ denotes the internal state of the $k$th particle. 

Since, by assumption, the internal degrees of freedom cannot be modified by the setup, the multiset of internal states that the particles are initially prepared in, $\{ \ket{\epsilon_1}, \dots \ket{\epsilon_N} \}$, remains invariant under scattering. Consequently, the first-quantized final state will always possess a \emph{permutation representation}, 
\eq \ket{\Phi^{1q}(\tilde g)}= \sum_{\sigma \in S_N} \bar g_{\sigma} \left(  \otimes_{j=1}^n \ket{\epsilon_{\sigma(j)}}_j \right) \label{permurepresent} ,\en
where $\bar g_\sigma$ is a coefficient that depends on the permutation $\sigma$, and all possible distributions of the single-particle states $\ket{\epsilon_j}$ on the spatial output modes are taken into account. By a parameter-counting argument, we see that not every state can be represented by the form (\ref{permurepresent}): The number of parameters (the number of $\bar g_\sigma$) is $N!$, whereas $N^N$ parameters are required for the description of a general state of $N$ particles each associated with an $N$-dimensional degree of freedom. For constant coefficients $\bar g_{\sigma}=\bar g$, the state (\ref{permurepresent}) is fully permutation-symmetric. In particular, any permutation-symmetric state with $d=2$ can be written in the \emph{Majorana-representation}  \cite{PhysRevA.83.042332,Majorana:1932qa}, i.e.~one can then always find a multiset $\{ \ket{\epsilon_j} \}$ to represent it in the form  (\ref{permurepresent}). Physically speaking, every permutation-symmetric state of qubits can be created with beam-splitters and phase-shifters \cite{PhysRevLett.102.053601}. 

In the present scattering scenario, the coefficients $\bar g_{\sigma}$ cannot be adjusted independently, but they are given as a function of the scattering matrix elements $V_{k,l}$, and they depend on the type of particles (bosons or fermions). Each component $\ket{\epsilon_{\sigma(1)}, \dots , \epsilon_{\sigma(N)}}$ in the state (\ref{permurepresent}) obtains the amplitude of the corresponding \emph{many-particle path}, for which the $\sigma(j)$th particle falls into the $j$th spatial output mode. Consequently, applying $\hat P_{1,b}$ to the state in (\ref{bigsum}) and identifying the coefficients $\bar g_{\sigma}$ in the equivalent representation (\ref{permurepresent}) yields 
\eq \bar g_{\sigma}= \text{sgn}_{\text{B/F}}(\sigma) \prod_{j=1}^N V_{\sigma(j),j} ,  \label{productrepresig} \en 
where $\text{sgn}_{\text{B/F}}(\sigma)$ denotes the signature of the permutation $\sigma$ for fermions, and is always unity in the case of bosons. That is to say, the final state for non-polarizing setups is a superposition of all possibilities to distribute the particles prepared at the $N$ spatial input modes over the $N$ spatial output modes. The amplitude for each of these permutations is given by the product of the scattering matrix elements as defined in (\ref{productrepresig}). \\

\subsection{Polarizing and polarization-manipulating setups}
\begin{figure}[ht]
\center
\includegraphics[width=\linewidth,angle=0]{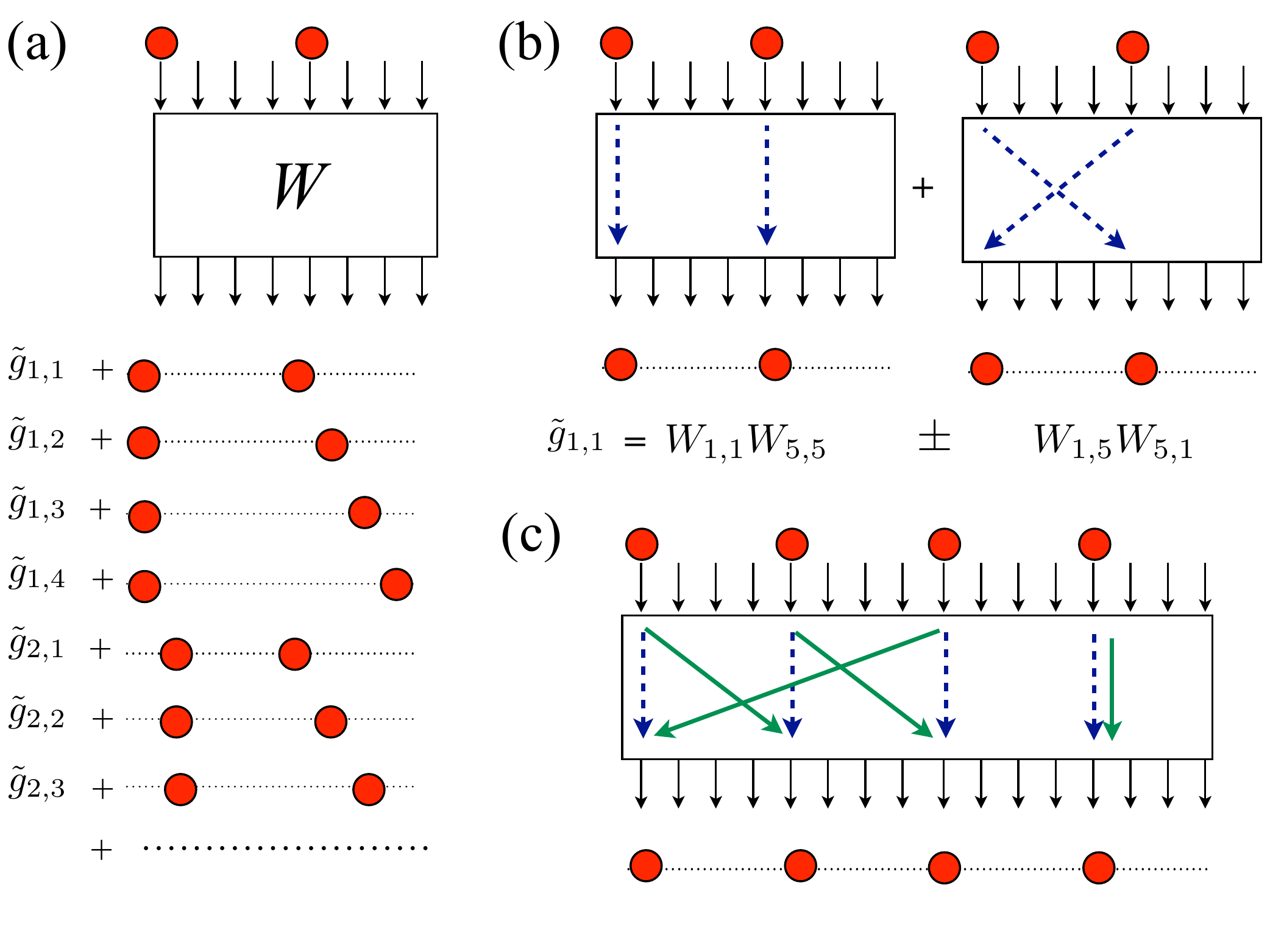} 
\caption{(color online)   Superposition of many-particle paths. (a) A state of $N=2$ particles that carry a $d=4$-dimensional degree of freedom is characterized by the coefficient tensor $\tilde g$, which indicates the amplitude of each basis state. (b) Each basis state (here: $\ket{1,1}$) is fed by the many-particle paths that contribute to the respective event. (c) In general, there are up to $N!$ of these paths, we show here two exemplary paths for $N=4$ particles.} \label{Superpositions}
\end{figure}

For polarizing setups, which do not manipulate the internal degrees of freedom of the particles, one cannot use the permutation representation to express the final state, since particles in distinct internal states can behave differently and the single-particle states $\{ \ket{\epsilon_k} \}$ are not preserved. We therefore treat this case together with polarization-manipulating setups.  In general, the target state's first-quantized notation is 
\eq \ket{\Phi^{1q}(\tilde g)} & =& \sum_{j_1=1}^d \dots \sum_{j_N=1}^d \tilde g_{j_1, \dots ,j_N} \ket{j_1, j_2 ,  \dots j_N} .\label{gep1}
 \en
For an initial state with all particles in the internal state $\ket{j_k=1}$ (remember that arbitrary local operations on the initial qudits can be absorbed in the scattering matrix $W$, as described in Section \ref{initialstateindepen},  we evaluate the coefficients $\tilde g_{j_1,\dots,j_N}$ via the application of $\hat P_{1,b}$ onto (\ref{bigsum}):
\eq 
 \label{gexpress}  \tilde g_{j_1, \dots ,j_N}    = \hspace{5cm} \\ \nonumber \sum_{\sigma \in S_N} \text{sgn}_{\text{B/F}}(\sigma)  \prod_{k=1}^N W_{d\cdot (\sigma[k]-1)+1,d(k-1)+j_k}  .
\en
For the interpretation of (\ref{gexpress}), we can apply an intuitive picture of many-particle paths \cite{1367-2630-14-9-093015}, which is also illustrated in Fig.~\ref{Superpositions}: Many distinct many-particle paths contribute to post-selected events with one particle per group of modes, and the entanglement in the final state originates from this superposition of distinct paths \cite{PROP:PROP201200079} (see Fig.~\ref{Superpositions} (a)). Each separable basis state $\ket{j_1, j_2 , \dots j_N}$ in the final state representation is fed by up to $N!$ distinct contributions (see (b) and (c)): The particle initially in mode $d\cdot (\sigma[k]-1)+1$ is redirected to mode $d\cdot(k-1)+j_k$, and the product of the corresponding amplitudes gives the many-particle amplitude that contributes to this specific scattering process. All these amplitudes are superimposed and interfere (see (b)), and their sum equals the coefficient $\tilde g_{j_1, \dots, j_N}$.

\subsection{Combinatorial bounds} \label{combibounds}
With the explicit state representations at hand (see (\ref{permurepresent}) and (\ref{gexpress})), we can now map the task ``given a certain infrastructure, can we create the state $\ket{\Phi^{1q}(h)}$ starting from a separable state?'' to a concrete mathematical problem: It corresponds to the  set of $d^N$ equations 
\eq \tilde g_{j_1\dots j_N}=\eta~ h_{j_1,\dots j_N}, \en for the matrix elements $W_{k,l}$, where the $\tilde  g_{j_1,\dots j_N}$ are given by (\ref{gexpress}) and $\eta$ accounts for the sub-normalization of the $\tilde g$. In general, this task is manifestly  rather difficult, since the coefficients $\tilde g_{j_1\dots j_N}$ are highly non-linear functions of the unknown parameters $W_{k,l}$. 

Still, the explicit representation (\ref{gexpress})  permits the identification of limitations that apply to entanglement generation via many-particle interference, in terms of an upper bound for the generalized Schmidt rank \cite{Eisert:2001uq}. 
\begin{widetext}
The bound can be derived from the representation of the final state $\ket{\Phi^{1q}(\tilde g)}$, which reads, according to (\ref{gep1},\ref{gexpress}), 
\eq 
\ket{\Phi^{1q}(\tilde g)}=\sum_{j_1=1}^d \dots \sum_{j_N=1}^d  \sum_{\sigma \in S_N} \text{sgn}_{B/F}(\sigma) \prod_{k=1}^N W_{d(\sigma[k] -1)+1, d(k-1) +j_k} \ket{j_1, \dots , j_N}  , \label{initstatecomb}
\en
and which can be re-written as
\eq
\ket{\Phi^{1q}(\tilde g)}= \sum_{\sigma \in S_N} \text{sgn}_{B/F}(\sigma)  \left[ \otimes_{m=1}^N  \left( \sum_{j_m=1}^d  W_{d(\sigma[m] -1)+1, d(m-1) +j_m} \ket{j_m}_m \right) \right] , \label{combreprese}
\en
where $\ket{j_m}_m$ denotes the state of the $m$th qudit prepared in the $j_m$th internal state. In other words, every state that is created by many-particle interference can be written as the sum of $N!$ (from the sum over all permutations $\sigma$) separable terms of the form (\ref{separablet}). 
\end{widetext}

This result immediately yields an upper bound for the Schmidt rank $R(\ket{\Phi^{1q}(\tilde g)})$ of the state $\ket{\Phi^{1q}(\tilde g)}$, 
\eq  R(\ket{\Phi^{1q}(\tilde g)}) \le N!  \label{rankbound} ~, \en
where the Schmidt rank is the minimal necessary number of summands in the representation of $\ket{\Phi^{1q}}$ as a sum of separable states~\cite{Eisert:2001uq}. 
 This fundamental bound for multipartite entangled states that are generated via many-particle interference can be understood via an intuitive combinatorial argument: Each possibility to distribute the particles among the output modes can only contribute one term in the representation of the emerging state, for bosons in the very same way as for fermions. Thereby, we have established a link between the number of many-particle paths and an entanglement measure, namely the generalized Schmidt number, which is defined as $\text{log}_2 R(\ket{\Phi^{1q}(\tilde g)}) $  \cite{Eisert:2001uq}.

\subsection{Initial states with multiple occupation} \label{multioccupinistate} 
Through our reference to (\ref{gexpress}), we assumed in Section \ref{combibounds} above that  the initial state is the $N$-qudit-state $\ket{\Phi^{1q}}=\ket{1,\dots,1}$, i.e.~a state with exactly one particle in each group of $d$ modes. If an initial physical state with larger occupancy is used (which does not admit an interpretation as $N$-qudit-state and which is only admissible for bosons), a tighter bound emerges. In this case, we can formally use the same initial state $\ket{\Phi^{1q}}$, but we account for the multiple occupation by equating the corresponding rows of the scattering matrix (see Section \ref{initialstateindepen}). When $r_j$ is the occupation number of each mode ($\sum_{j=1}^n r_j= N$), the scattering matrix $W$ effectively reflects this multiple occupation by possessing  $r_j$ equal rows.  The outer sum in (\ref{combreprese}) reduces to only 
$ N!/(\prod_{j=1}^n r_j! ) $ terms, since for each $\sigma \in S_N$, there are $\prod_{j=1}^n r_j!$ permutations that leave the summand 
 \eq \otimes_{m=1}^N   \sum_{j_m=1}^d  W_{d(\sigma[m] -1)+1, d(m-1) +j_m} \ket{j_m}_m  \en  invariant. This can, again, be interpreted in a picture of many-particle paths: When several particles are initially prepared in the very same state, a permutation of these particles does not lead to a distinct many-particle amplitude. Therefore, the number of distinct many-particle amplitudes is reduced by a factor $\prod_{j=1}^n r_j!$.

As an extreme example, it is not possible to extract entanglement from an ideal Bose-Einstein-Condensate (BEC), which is described by 
\eq \ket{\Psi^{2q}_{\text{BEC}}}=\frac{ \left( \hat a^\dagger_1 \right)^N } { \sqrt{ N! } }  \ket{\text{vac}} , \label{beclikestate} \en
i.e.~a state of $N$ particles with all particles in the same mode, since $N!/(\prod_{j=1}^n r_j!)=1$ for $\vec r=(N,0,\dots,0)$. Post-selected events with one particle in each output mode can only be reached via one many-particle path, such that the Schmidt rank of the resulting state is always unity. This generalizes the result of \cite{Sasaki:2011uq}, where it was found that no entanglement can be extracted from a BEC when no manipulations of the internal state dynamics are allowed. One consequence of our results is that even such manipulation on the internal degrees of freedom is of no help.

\subsection{Entangled initial states}\label{concatenationsec}
In turn, the use of an entangled initial state allows one to increase the upper bound on the Schmidt rank of the final state: The combinatorial bound applies to every single separable term in the representation of the initial state in its generalized Schmidt form. Given an initial  Schmidt rank $R_{\text{in}}$, the combinatorial bound is extended up to $R_{\text{in}} N!$. In other words, the concatenation of many-particle interference processes and post-selection can be used to increase the Schmidt rank step by step-wise, by a factor of up to $N!$ in each step. 

\subsection{Interpretation}
The combinatorial bound should not be regarded as a purely quantum phenomenon: It also applies to a classical probabilistic process in which $N$ distinguishable objects are randomly distributed to $N$ observers. The number of distinct outcomes that the $N$ observers can experience is, again, bound by $N!$, just like the generalized Schmidt rank is according to (\ref{rankbound}). If some of the objects are identical, permutations that exchange these objects do not lead to inequivalent events, just like the multiple initial occupation of any mode reduces the rank, as discussed in Section \ref{multioccupinistate}. The quantum properties of bosons and fermions, i.e.~their (anti)-commutation relations which provide an alternating sign for fermions in the sums (\ref{bigsum},\ref{gexpress}), do not play any role, and the bound applies equally well to both species. It is, furthermore, independent of the presence of polarizing components. In contrast, we will see that bosons and fermions indeed do hold a different potential for entanglement generation, as will be shown by means of the dimensionality of the manifold of accessible states in the next Section.

\section{Dimension of the manifold of accessible states} \label{Dimsetsse}
For systems of large dimensions $d\gtrsim N$, the combinatorial bound (\ref{rankbound}) dominates the set of accessible states. 
By a parameter-counting argument, however, we see that not all states can be created when the number of particles $N$ is large, for \emph{any} dimension $d \ge 2$: On the one hand, the number of complex parameters for a general $N$-partite quantum state with a $d$-dimensional internal degree of freedom is $d^N$ (including normalization and global phase). On the other hand, the number of complex parameters of the scattering matrix $W$ that controls the final state (\ref{initstatecomb}) is only $d \cdot N^2$: We are given the initial state $\ket{\Phi^{1q}}=\ket{1,1,\dots,1}$, and only the matrix elements $W_{d(k-1) +1, j}$ with $1 \le k \le N$ and $1 \le j \le d N$ appear in the sum (\ref{initstatecomb}). The exponential growth of the parameters of entangled states with the number of involved particles reflects the complexity of many-particle quantum mechanics, which cannot be reproduced by the limited amount of free parameters in the interference-based setup under investigation. 

\subsection{Fermionic and bosonic bounds}
The rough upper bound provided by the above parameter-counting argument can be ameliorated by studying the structure of the state coefficients in (\ref{initstatecomb}) in more detail. This allows us to obtain a tight upper bound for the dimensionality of the manifold of accessible states. We denote by  $\Xi_{\text{F(B)}}$ the manifold of states of the form (\ref{initstatecomb}) for fermions (bosons), which is spanned by all matrices $W$. In contrast to the combinatorial bound, we find a striking difference between bosons and fermions. The derivation of the bounds is presented in Appendix \ref{appdimbounds}. For fermions, we find 
\eq 
\text{dim}(\Xi_{\text{F}}) \le (d-1)N^2-N+2  \label{boundfermsd} ,
\en
while for bosons, we have 
\eq
\text{dim}(\Xi_{\text{B}}) \le d N^2 -2N+2 \label{boundbosons} .
\en
Roughly speaking, the computation of scattering amplitudes for fermions involves the evaluation of determinants, which is facilitated by the symmetry properties of this function, in particular, by the property $\text{det}(AB)=\text{det}(A)\text{det}(B)$. Scattering amplitudes for bosons instead involve the \emph{permanent} \cite{Ryser:1963oa,Sachkov:2002fk,Aaronson:2010fk}, a complex-valued function of matrices similar to the determinant, but that does not feature equivalent symmetry properties. The number of effective parameters for bosons is therefore larger than for fermions.

The bosonic bound (\ref{boundbosons}) is only applicable when it is smaller or equal to the dimension $d^N$ of the total state space. For $N=2$, and for $d=2, N=3,4$, the bound exceeds the total dimensionality of the many-particle Hilbert-space, $N^d$; it is thus not relevant. For large systems, the bound strongly limits the manifold of accessible states, as we will also see in the next Section in Table \ref{lowerboundsfbtt}. 

\subsection{Tightness of bounds} \label{tightness}
The established inequalities (\ref{boundfermsd}) and (\ref{boundbosons}) provide upper bounds to the dimensionality of the manifold of entangled states that can be created by fermions and bosons, respectively. For small dimensions $d$ and small particle numbers $N$, we  found numerically that the equality holds in (\ref{boundfermsd}) and (\ref{boundbosons}). One thus indeed possesses a number of independent parameters that corresponds to the right-hand side of (\ref{boundfermsd}), (\ref{boundbosons}). For this purpose, we evaluated the rank of the Jacobian \eq \frac{\partial g_{j_1, \dots, j_N}} {\partial W^\prime_{l,k}}  ,  \label{jacobian} \en for random scattering matrices $W$. The results are shown in Table \ref{lowerboundsfbtt}: they always match the encountered upper bound. 

\begin{table}[ht] \center
\begin{tabular}{r|rrrrrrr|rrrrrr}
 & \multicolumn{4}{l}{ $N$  Bosons } && && \multicolumn{6}{l}{ $N$  Fermions }\\
             & $2$ & $3$ & 4 & 5 & 6 & 7 &8& 2 & 3 & 4 & 5 & 6 & 7    \\  \hline
$d=2$ & {\color{blue}{\underline{4}}} & {\color{blue}{\underline{8}}} & {\color{blue}{\underline{16}}} & {\color{blue}{\underline{32}}}&  62&  86 & 114 & 
{\color{blue}{\underline{4}}} & {\color{blue}{\underline{8}}} & 14 & 22 & 32 & 44   \\
$3$ & {\color{red} (8)}  &  23  &  42&  67  &  98  &  & & 
8 & 17& 30 & 47 & 68  \\
$4$ & {\color{red} (12)}    &   32	  &   58 	&   92 	& 	134	& 	& &  
12 & 26 & 46 & 72 \\
$5$ &  {\color{red} (16)} 	& 41  & 74	& 117 &	 & 	&  &  
16 & 35 & 62&  97 \\ 
\end{tabular} 
\caption{(color online)  Numerically obtained lower bounds for the dimensionality  $\text{dim}(\Xi_{\text{B/F}})$ of the accessible set of states, for unconstrained setups, according to Eq.~(\ref{jacobian}). Blue {\underline{underlined}} numbers indicate combinations of $d$ and $N$ which are not subject to any constraint, i.e.~all entangled states can be created via many-particle interference. Red numbers (in parenthesis) refer to combinations of $(d,N)$ for which the combinatorial bound (\ref{rankbound}) is tighter than the bosonic bound (\ref{boundbosons}). All numerically inferred lower bounds match the analytically obtained upper bounds (\ref{boundfermsd}) and (\ref{boundbosons}) or the maximal dimensionality that can be attained according to the combinatorial bound  (\ref{rankbound}).  } \label{lowerboundsfbtt}
\end{table}

\subsection{Interpretation and consequences} \label{disconstrrrr}
Our results on the effective number of parameters for bosons and fermions demonstrate that the combinatorial bound (\ref{rankbound}) is not the only restriction to the set of accessible states.  The complexity of general multi-partite entangled states cannot be reproduced in interference-based setups, simply due to the limited number of parameters. When fermions are used, the symmetry properties of the determinant reduce the effective number of parameters in comparison to bosons. Indeed, the computation of fermionic many-particle scattering amplitudes is computationally much less demanding than the corresponding task for bosons \cite{Aaronson:2010fk}, and this computational complexity is reflected in the larger manifold for bosons.

This larger manifold of states that are accessible for bosons comes at an expense: For fermions, the entanglement generation process is assisted by the Pauli principle, which forbids multiply occupied modes that correspond to unsuccessful events. The average tendency of bosons to bunch, on the other hand, enhances unsuccessful events. For larger dimensions $d$, the total number of modes $n$ increases, and the Pauli principle becomes less dominating, as well as the bunching tendency of bosons is less prominent: Due to the lower particle density, multiply occupied states become less probable in general, since many more possible states with at most one particle in each of the $n$ modes emerge. Simultaneously, the quotient of the bounds to the manifold dimension of fermions to the manifold dimension of bosons, $\text{dim}(\Xi_{\text{F}})/\text{dim}(\Xi_{\text{B}})$  (see (\ref{boundfermsd}) and (\ref{boundbosons})), converges to unity for $d \rightarrow \infty$, i.e.~the advantage of bosons with respect to fermions regarding the dimensionality is jeopardized. In other words, there is a trade-off between success probability and the dimensionality of the accessible manifold. 

Given the dimensionality estimate (\ref{boundbosons}) for qubits that are carried by bosons, we can expect that all four-partite and five-partite qubit states can be created with linear optics, with experiments similar to \cite{Wieczorek:2009ff,Wieczorek:2008uq,Wieczorek:2009fk,Wieczorek:2009fe,Prevedel:2009ec}. For six photons, we find that only 62 of a total of 64 dimensions are spanned. Given the manifold possibilities to create entangled six-photon states by parametric down-conversion processes, it is, however, possible that one may find ways to create the unaccessible states starting from entangled initial states. 

Significant progress has been achieved in the design of alternative, efficient single-photon sources \cite{Benson:2000db,Santori:2002qf,silberhornneu}. However, our results indicate that extending experiments to larger particle numbers is not only limited by technological issues, but fundamental limitations to the accessible states will severely come into play for large particle-numbers. For example, for $N=8, d=2$, $\Xi_{\text{B}}$ has only 114 dimensions, significantly less than the dimension of the Hilbert space, 256. In analogous experiments with fermions, e.g.~with cold atoms in optical lattices, one will experience a bound for the accessible states already for four particles. 

For bosons, the inclusion of loss in the setup, i.e.~the use of non-unitary scattering matrices $W$, permits the generation of states that are inaccessible for unitary matrices: Consider a lossless setup described by the unitary $d N \times d N $ matrix $W$, i.e.~the matrices $X,Y,Z$ in the representation (\ref{embedma}) are empty. The number of parameters that affect the relevant $dN \times N$-dimensional sub-matrix $W^\prime$ of $W$  can be inferred by counting the Euler-angles of $W$ that do not affect $W'$. The total number of complex Euler-angles that define $W$ amounts to $N d (Nd-1)$, while $N(d-1) (N (d-1)-1)$ of these are irrelevant for $W'$. The difference yields the number of parameters $K$, \eq K=\left(d-\frac 1 2 \right)N^2-\frac N 2  .\en This value is exceeded by the dimensionality of bosons, $\text{dim}(\Xi_{\text{B}})$ (\ref{boundbosons}). The inclusion of lossy setups (or, equivalently, of the multiple population of modes in the initial state, see Section \ref{multioccupinistate}) thus offers advantages over lossless, unitary setups. For fermions, however, the number of parameters of a unitary matrix $U$ is larger than our dimensional bound, $\text{dim}(U)>\text{dim}(\Xi_{\text{F}})$, which suggests that  no advantage is accomplished for fermions by including loss in the setup.

\section{Conclusions and outlook} \label{concluout}
The emergence of quantum correlations from the propagation and detection of bosons and fermions can be described by the presented  formalism, which also provides an intuitive interpretation: Particles are re-distributed from input modes  to output modes, and the probability $p_{j,k}=|W_{j,k}|^2$ to fall into each output mode $k$ depends on the initial state of the particle, $j$. This process has a purely classical analogy: The random distribution of distinguishable objects to parties/observers also leads to classical correlations between the objects that are subsequently observed by the parties. The combinatorial bound can then be understood from a rather intuitive, classical perspective: The number of possibilities to redistribute objects among observers is limited. Here, we have considered the fully coherent version of this process, and the resulting correlations are of quantum nature, but the combinatorial bound remains valid. It remains to be studied how a partial loss of indistinguishability affects the quality of the emerging final state, which requires a framework in which the deteriorating distinguishing degree of freedom is incorporated explicitly \cite{PROP:PROP201200079,tichydiss,TichyLim}.

The physical process of \emph{redistributing} particles among observing parties is the very basis for a panoply of experiments and theoretical schemes. We aimed at providing a common framework for the creation and detection of entangled states via the deletion of which-way information, and obtained an explicit representation (\ref{gexpress}) of all states that can be created or detected by linear optics setups. Depending on the experimental capabilities, i.e.~whether the device that is equivalent to a polarizing beam-splitter can be implemented or not, the form of the scattering matrix is very different, and different types of states can become accessible. The combinatorial bound, however, cannot be circumvented by increasing the complexity of the scattering setup. The computational complexity of the permanent and determinant, which appear in the scattering amplitudes for bosons and fermions, respectively, is reflected by the effective number of parameters that are available to generate entangled states. In contrast to the combinatorial bound, this dimensional bound can only be understood in a quantum-mechanical way. It reveals that the intrinsic power of photons for entanglement generation lies in the computational complexity of many-boson processes.

In general, any experiment in which particle-carried entanglement is created without direct interaction between the constituents can be described by a scattering matrix $W$, and the resulting state has a representation of the form (\ref{gexpress}). Given the experimental limitations, one can thus optimize the creation of entanglement by finding the optimal matrix $W$. 

Some proposals that encode qudits in a different way than we have assumed here can circumvent the combinatorial bound, and can reach states that are not represented by (\ref{gexpress}). For example, mode-entanglement \cite{Lee:2002fk,Pryde:2003fk,Brougham:2011vn}, i.e.~entanglement in the particle-number degree of freedom, is not subject to the bound, as well as schemes in which qudits are carried by several particles instead of a single one \cite{Baek:2007mi,PhysRevA.78.042321,Halevy:2011cr,Lanyon:2008bh} or  techniques that exploit ancilla-particles \cite{Bouwmeester:1999ys,Ikuta:2011ys}. An extension of our framework that explicitly incorporates these methods may shade light on their potential and quantify their advantage with respect to the paradigm that we considered here. We expect that the here-encountered advantage of bosons over fermions will also persist in these extended scenarios, based on the computational complexity of the scattering of bosons.

The prospect for many-particle entanglement creation with photons is restricted by the dramatic decrease of the success rate for increasing particle number $N$, since unsuccessful events with multiply occupied spatial output modes then become more and more probable, although there are approaches to tackle this inefficiency by using the time-bin as an additional discriminating degree of freedom \cite{Megidish:2012uq}. The very understanding of multipartite entanglement being a widely open issue, the state representation (\ref{gexpress}) offers a way to understand at least the structure of the states that can be created via the exploitation of the indistinguishability of the particles.  In particular, completely anti-symmetric and completely symmetric states are contained in the set of  states achievable for fermions and bosons, respectively (see Eq.~(\ref{fullsymmmm})). Since these special classes are rather well understood \cite{1367-2630-12-7-073025,Bastin:2009ye,Mathonet:2010fu,PhysRevA.77.012104,PhysRevA.83.042332,PhysRevLett.106.180502}, the hope is fed that one may characterize the more general state (\ref{gexpress}) via entanglement measures or via its non-local properties, and thereby also characterize the potential of non-linear single-photon schemes \cite{Peyronel:2012fk}.

In particular, the question arises whether the manifold of accessible states for fermions is a \emph{submanifold} of the accessible states for bosons. This appears suggestive from the clear hierarchy of the bounds (\ref{boundfermsd},\ref{boundbosons}) for the dimensionalities that we have encountered, but, in turn, also questionable, since fully symmetric and fully anti-symmetric states, which are accessible for  bosons and fermions, respectively, possess very distinct entanglement properties \cite{PhysRevA.77.012104}.   
Related to that, it is also widely open whether states that are accessible for bosons but inaccessible for fermions feature particular physical properties. Furthermore, the influence of restrictions on the setup, i.e.~the use of non-polarizing setups instead of the unconstrained form, remains to be studied. The answers to these questions may contribute further to an understanding of the structure of multipartite entangled states.

\acknowledgements 
We would like to thank Hyang-Tag Lim and Young-Sik Ra for enlightening discussions.  M.C.T. acknowledges support by the German National Academic Foundation and by the Alexander von Humboldt-Foundation through a Feodor Lynen Fellowship, A.B. acknowledges partial support through the EU-COST Action MP1006 ``Fundamental Problems in Quantum Physics'' and through DFG Research Unit 760. 

\appendix

\section{Accommodation of existing schemes} \label{classification}
Many existing schemes that seem rather unrelated to each other at first sight can be put onto a common footing. In the following, we give an exemplary overview over various methods that can be described by the formalism described in this article, which stresses the shared underlying physical roots of the different theoretical and experimental methods.

\subsection{Theoretical proposals}
\subsubsection{Scattering on the Fourier multiport}
Several multipartite entangled states can be created with Fourier multiport devices via the scattering of bosons that are prepared in certain initial states \cite{Lim:2005bf}. The setup is non-polarizing, with 
\eq W=V\otimes \mathbbm{1}_2, \  \text{ and } \ V=U^{\text{Fou}}, \en where $\left( U^{\text{Fou}} \right)_{j,k}=e^{i \frac{2 \pi}{N} (j-1)(k-1)}$ is the Fourier matrix. No devices that are equivalent to polarizers need to be used, such that the infrastructure is independent of the specific degree of freedom that is entangled. The final state reads (as given in (\ref{permurepresent})) 
\eq \ket{\Phi^{1q}(\tilde g)}=\sum_{\sigma \in S_N} \left(  \prod_{j=1}^N U^{\text{Fou}}_{\sigma(j),j}  \right) \otimes_{k=1}^N \ket{\epsilon_{\sigma(k)}}_k , \en
where the $\ket{\epsilon_k}$ are the initial internal states of the bosons at the input modes. In particular, a W-type state \cite{Duerr:2000pj} can be created by preparing one photon in horizontal, $\ket{H}$,  and the remaining $N-1$ photons in vertical polarization, $\ket{V}$. Due to the cyclic invariance of the Fourier matrix, it is ensured that all  amplitudes of the distinct components with the $j$th particle in horizontal polarization ($\ket{H,V,V,V,\dots }$, $\ket{V,H,V,V,\dots }$, etc.) enter with the same amplitude. For four particles, a GHZ-state of four photons can equally well be produced with the initial state $\ket{V,V,H,H}$. 

\subsubsection{Free space propagation}
A free-space photon-propagation scheme \cite{PhysRevLett.99.193602,Maser:2009pi,Schilling:2009fk,PhysRevLett.102.053601,Maser:2010fu} allows to generate a variety of entangled states. Both, entanglement projection (with consequent  entanglement swapping) \cite{Schilling:2009fk}, and entanglement creation \cite{Maser:2010fu} are possible. In all cases, photons propagate in free space, and they are collected by an array of $N$ detectors, each of which projects the incoming photon onto a certain polarization state. Each photon has the same probability to be absorbed by each detector. 
In particular, Dicke states, i.e.~states $\ket{S,m}$ which are simultaneous eigenstates of the total squared spin operator $\hat S^2$ and its $z$-component $\hat S_z$, with eigenvalues $S$ and $m$, respectively, are accessible \cite{PhysRevLett.99.193602}. 
 More generally, all permutation-symmetric states can be created \cite{Maser:2010fu}, and -- with few modifications of the scattering matrix -- all $2^N$ distinct total angular momentum eigenstates \cite{Maser:2009pi,zanthierunpubl}.

The scattering matrix $W$ is non-polarizing, and its submatrix $V$ (see (\ref{nonpolmatrix})) is the constant matrix with $V_{j,k}=\sqrt{p}$, where $p$ is the probability that a photon is registered by the detector. For the state representation, we can use (\ref{permurepresent}) to find 
\eq \ket{\Phi^{1q}_{\text{fin}}}&=& \sum_{\sigma \in S_N} \left(\prod_{j=1}^N V_{\sigma(j),j} \right) \otimes_{k=1}^N \ket{\epsilon_{\sigma(k)}}_k \nonumber \\ &=& \sqrt{p^N} \sum_{\sigma \in S_N}  \otimes_{k=1}^N \ket{\epsilon_{\sigma(k)}}_k \label{fullsymmmm} .\en
In other words, a fully permutation-symmetric state emerges \cite{1367-2630-12-7-073025,Bastin:2009ye,PhysRevLett.106.180502,Mathonet:2010fu,PhysRevA.83.042332}. 

\subsection{Experimental implementations}
Many qualitatively distinct classes of entangled states \cite{Lu:2009zr,Eibl:2003ly,Lu:2007ve,Walther:2005qf} have been realized in experiments that employ photons as carriers of entanglement. In order to exemplify the versatility of many-photon entanglement generation, we discuss in more detail the projection onto a tripartite entangled GHZ-state \cite{Lu:2009fk} and the versatile tunability of distinct SLOCC classes for four-photon states \cite{Wieczorek:2008uq}. 

\subsubsection{GHZ entanglement swapping} 
\begin{figure}[ht]
\center
\includegraphics[width=\linewidth,angle=0]{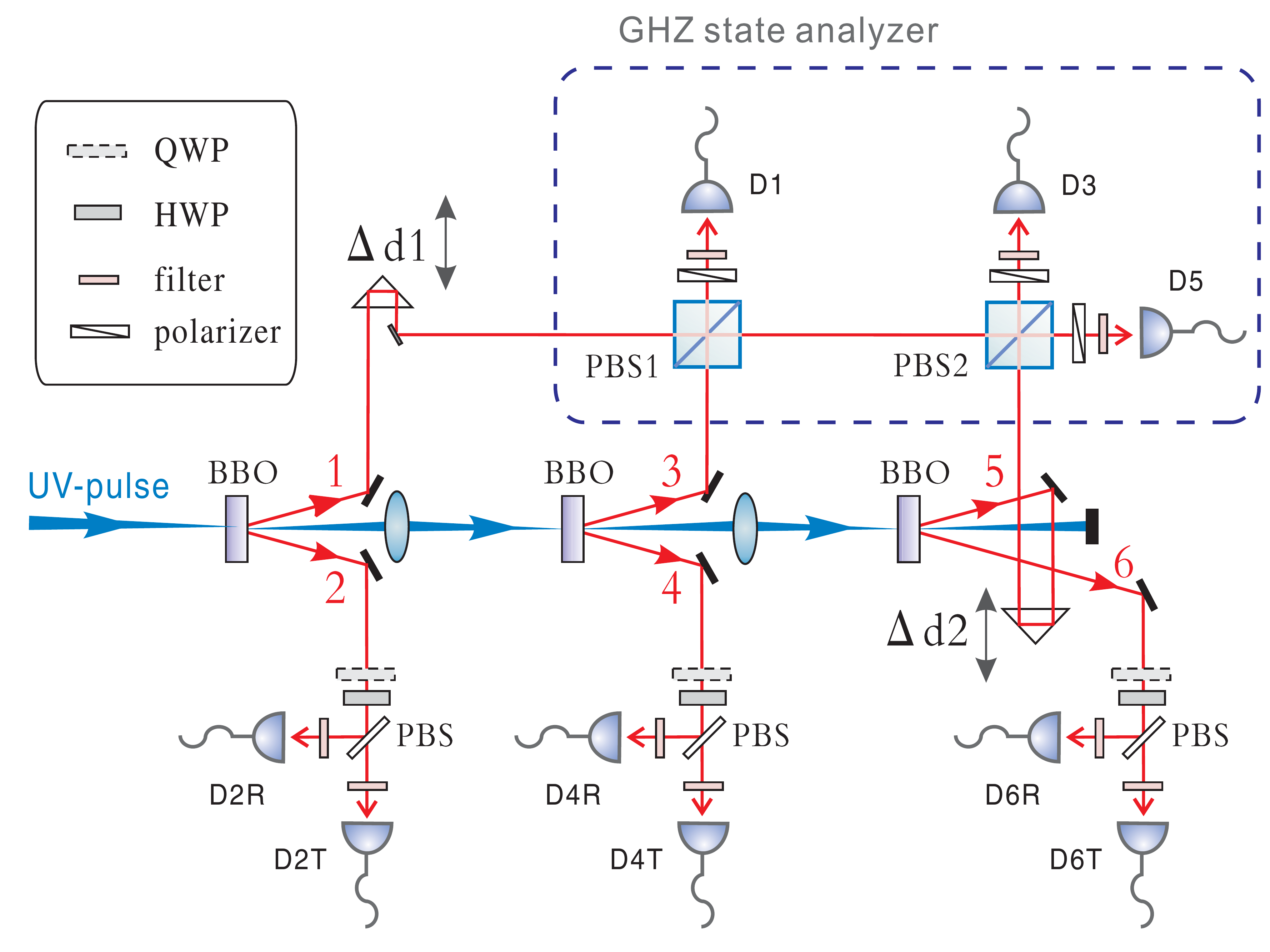} 
\caption{(color online)   GHZ entanglement projection.  The photon pairs (1,2), (3,4) and (5,6) are initially polarization-entangled, as described by Eq.~(\ref{suphi}). By the projection of photons 1, 3 and 5 onto the GHZ-state (at detectors D1, D3, D5, by the implementation of the scattering matrix (\ref{ghz3scat})), photons 2, 4, and 6 are also left in the GHZ-state. Courtesy of C.-Y.~Lu, T.~Yang and J.-W.~Pan \cite{Lu:2009fk}.} \label{Panfigure}
\end{figure}

Entanglement swapping among three parties was experimentally demonstrated \cite{Lu:2009fk} with the setup reproduced in Fig.~\ref{Panfigure}. A laser pulse that induces parametric down-conversion on three aligned crystals (BBO in the Figure) results in the creation of 
\eq \ket{\Phi^{1q}_{\text{ini}}} = \ket{\Phi^+}_{1,2} \otimes \ket{\Phi^+}_{3,4} \otimes \ket{\Phi^+}_{5,6} , \label{suphi} \en
where $\ket{\Phi^+}_{k,l}$ denotes the $\ket{\Phi^+}$ Bell-state in the polarization degree of freedom, shared between the photons in the spatial modes $k$ and $l$: 
\eq 
\ket{\Phi^{+}}_{k,l}= \frac{1}{\sqrt{2}} \left(  \ket{H,H}_{k,l} + \ket{V,V}_{k,l} \right) .
\en 
Photons 1, 3 and 5 are then projected onto a GHZ-state, \eq \ket{\text{GHZ}}_{1,3,5}=\frac{1}{\sqrt{2}} \left( \ket{H,H,H} + \ket{V,V,V}  \right) ,\en via a GHZ-state analyzer \cite{Pan:1998fk} (dashed area in Fig.~\ref{Panfigure}). The polarizing beam-splitters (PBS) reflect vertically polarized photons and transmit horizontally polarized ones, while the polarizers after the PBS only transmit the $\ket{+}=(\ket{H}+\ket{V})/\sqrt{2}$-component of the light field. The resulting scattering matrix $W_{\text{GHZ}}$ that describes the propagation of the photons in the input modes 
\eq \ket{H}_1, \ket{V}_1, \ket{H}_3, \ket{V}_3, \ket{H}_5, \ket{V}_5 ,\en
to the detectors  $D1$, $D3$, $D5$ reads 
\eq  
W_{\text{GHZ}} = \frac 1 {\sqrt 2} \left(
\begin{array}{cccccc}
 0 & 0 & 1  \\
 1 & 0 & 0 \\
 1 & 0 & 0 \\
 0 & 1 & 0 \\
 0 & 1 &0 \\
 0 & 0 & 1 
\end{array}
\right) \label{ghz3scat}
 ,\en
i.e.~it corresponds to a combined action of polarizing beam-splitters and polarizers, as described in Section \ref{polasetups}. The state at modes 2,4 and 6 is then measured in different bases to verify quantum correlations. The path lengths $\Delta d1$ and $\Delta d2$ (see Figure) are adjusted by monitoring four-photon and six-photon coincident events. As discussed in Section \ref{CreationDetection}, the matrix $W_{\text{GHZ}}^\dagger$ can also be used for the \emph{creation} of a GHZ-state.

\subsubsection{Family of four-photon states}
The setup for the projection onto the GHZ state discussed above projects onto one very specific state. One can, however, also scan through entire families of entangled states by tuning the parameters of the scattering matrix $W$. This was realized in a four-photon experiment \cite{Wieczorek:2008uq,Wieczorek:2009fk}, in which one starts with a four-photon state created via parametric down-conversion, which reads 
\eq \ket{\Psi^{2q}_{\text{ini}}}&=& \frac{1}{2 \sqrt 3} \left( \left(\hat a^\dagger_{1,H} \hat a^\dagger_{2,V}\right)^2  + \left(\hat a^\dagger_{1,V} \hat a^\dagger_{2,H}\right)^2 \right. \nonumber \\ && \left. + 2 \hat a_{1,H}^\dagger\hat a_{1,V}^\dagger \hat a_{2,H}^\dagger\hat a_{2,V}^\dagger \right) \ket{\text{vac}}, \label{wit4tune} \en
i.e.~there are always two photons of horizontal and of vertical polarization, respectively, and there is one component with the photons distributed among all four physical modes and two components with two photons in the same mode. 

\begin{figure}[ht]
\center
\includegraphics[width=.8\linewidth,angle=0]{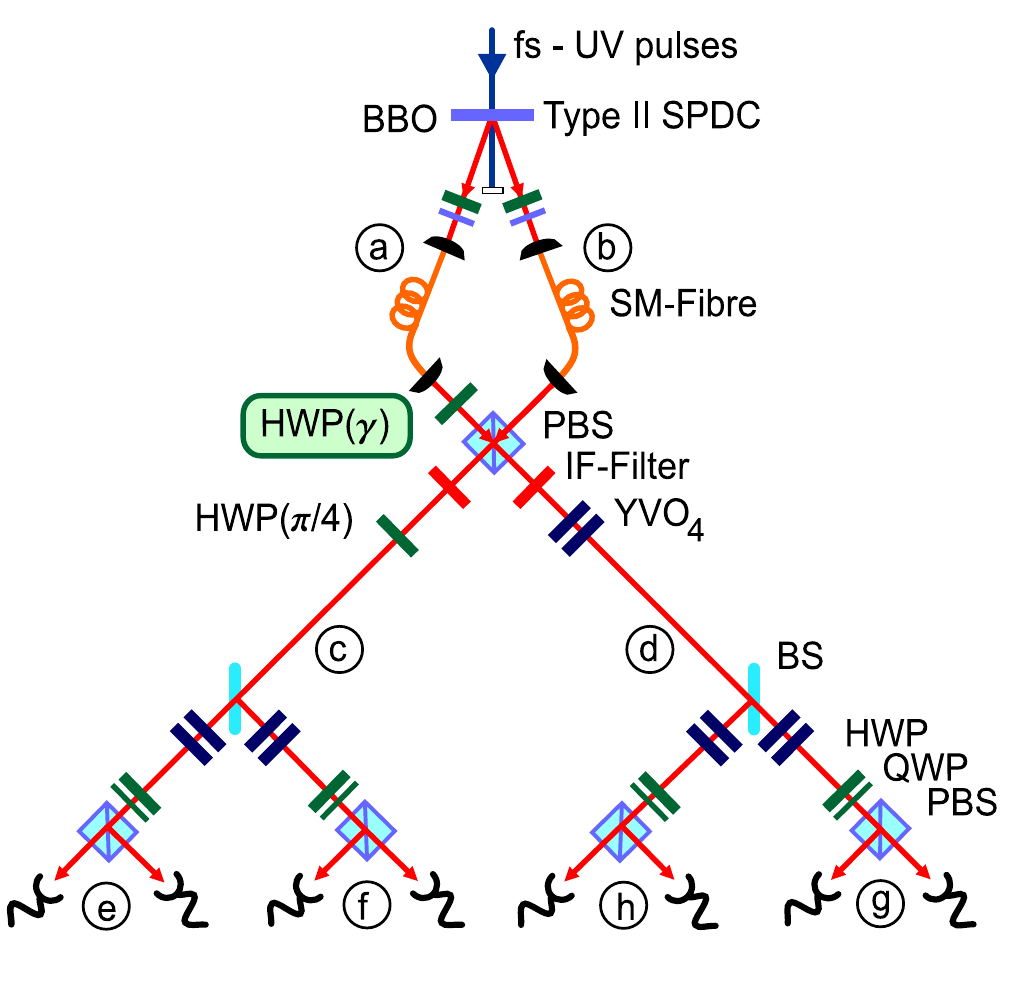} 
\caption{(color online)   Creation of a family of entangled four-photon states. Photons are created via spontaneous parametric down-conversion (SPDC), and injected into single mode (SM) fibres. Two input modes (a) and (b) are combined at the PBS and filtered by interference filters (IF). The resulting modes (c) and (d) are, subsequently, split into four modes (e), (f), (g), and (h) by two additional beam splitters (BSs). The initial state (\ref{wit4tune}) is thus scattered on a setup which implements (\ref{witlefscatmat4}). The final quantum-information state (\ref{finalstatewit4}) depends on the polarization-rotation parameter $\gamma$. Courtesy of W.~Wieczorek, C.~Schmid, N.~Kiesel, R.~Pohlner, O.~G\"uhne and H.~Weinfurter \cite{Wieczorek:2008uq}.} \label{Witlefpic2}
\end{figure}

A variable polarization-rotation with angle $\gamma$ is applied to the spatial mode $\hat a_{1}^\dagger$ (see HWP($\gamma$) in Fig.~\ref{Witlefpic2}, we identify {\sf (a)} and {\sf (b)} in the figure with the first and second spatial input modes). The two spatial modes are then combined at a polarizing beam-splitter ({\sf PBS} in the figure) that reflects vertically polarized photons. The polarization in one of the outgoing modes ({\sf (c)} in the figure) is then flipped by a half-wave plate (HWP($\nicefrac{\pi}{4}$)), and the spatial modes are split into two parts each, such that there are four final modes which can carry two possible polarizations (in the figure: {\sf (e), (f), (g), (h)}). The scattering matrix that describes the evolution of the four distinct input modes ($1H,1V,2H,2V$) to the output modes ($1H,1V,$\dots$,4H,4V$) reads
\eq 
W=\frac{1}{\sqrt{2}} \left(
\begin{array}{cccccccc}
i s & 0 & i s & 0 & c & 0 & c & 0 \\
i c & 0 & i c & 0 & -s & 0 & -s & 0 \\
0 & -1 & 0 & -1& 0 & 0 & 0 & 0 \\
0 & 0 & 0 & 0 & 0 & i & 0 & i 
\end{array}
 \right)
, \label{witlefscatmat4}
\en 
where $s=\text{sin}(2 \gamma)$ and $c=\text{cos}(2 \gamma)$, i.e.~the setup manipulates the polarization, and it depends on the polarization rotation angle $\gamma$. 
The result of a post-selected event with one particle in each output mode is a final state of the form 
\eq \ket{\tilde \Phi^{1q}} = \hspace{4.5cm}  \\
\nonumber \frac{1}{2 \sqrt 3} \left( \sqrt{2} \sin^2(2\gamma) \ket{\text{GHZ}_4} + \cos(4 \gamma) \ket{\Psi^+}_{1,2}  \ket{\Psi^+}_{3,4} \right),  \label{finalstatewit4} \en
i.e.~it is a coherent superposition of a bi-separable double-Bell state and a four-particle GHZ-state, where \eq 
\ket{\Psi^+}_{k,l}&=&\frac{1}{\sqrt{2}} \left( \ket{H,V}_{k,l} + \ket{V,H}_{k,l}  \right) , \\
\ket{\text{GHZ}_4}&=&\frac 1 {\sqrt 2}\left( \ket{H,H,V,V} +  \ket{V,V,H,H} \right). \en The success probability is $\left( \cos (4 \gamma)^2 + 2 \sin (2 \gamma)^4 \right)/12 $. The parameter $\gamma$ allows one to tune through several SLOCC-inequivalent entangled states of four photons \cite{PhysRevA.65.052112}. 

\section{Derivation of bounds on the dimensionality} \label{appdimbounds}
We derive here the upper bounds (\ref{boundfermsd}), (\ref{boundbosons}) for the dimensionality $\text{dim}(\Xi_{\text{F}})$ and $\text{dim}(\Xi_{\text{B}})$ of the manifold of accessible states, respectively. For this purpose, we regroup the matrix elements of $W$ that appear in the state representation (\ref{gexpress}) in a smaller matrix $W^\prime$ of dimension $N\times d N$: \eq W^\prime_{k,j}=W_{(k-1)d +1, j} , \en
such that $W^\prime$ contains all relevant parameters (remember from Section \ref{localunitaries} that we can always assume the initial separable state $\ket{\Phi^{1q}}=\ket{1,\dots, 1}$). Formally speaking, in (\ref{initstatecomb}) the $d N^2$ elements of the scattering matrix $W^\prime_{j,k}$ are mapped onto the $d^N$-dimensional space of the $\tilde g_{j_1,\dots j_N}$. The dimensionality of the accessible manifold $\Xi_{{\text{F/B}}}$ is the effective number of  complex parameters that are necessary to describe the $\tilde g_{j_1,\dots j_N}$.

\subsection{Fermionic qubits}
The dimensionality of the accessible space for fermions, $\text{dim}(\Xi_{\text{F}})$, follows from mathematical properties of the determinant, since the coefficient $\tilde g_{j_1, \dots, j_N}$ in (\ref{gexpress}) can be expressed as the determinant of sub-matrices of $W^\prime$. 

\label{fermionicqubits}
We start with the case of qubits for simplicity, and re-write $W^\prime$ as a vector of $2N$, $N$-dimensional column vectors,
\eq
W^\prime =\left( \vec w_1 , \vec w_2 , \dots , \vec w_{2 N} \right) .
\en
With (\ref{gexpress}), we can write each coefficient of the state representation, $\tilde g_{j_1, \dots, j_N}$, as the determinant of the square matrix that consists of the respective columns of $W^\prime$, i.e. 
\eq 
\tilde g_{j_1,\dots j_N} = \text{Det}\left[  \left( \vec w_{j_1}, \vec w_{2+j_2}, \dots \vec w_{2N-2+j_N}\right) \right] . \label{initialrepredet}
\en
If $W^\prime$ does not have full rank ($N$), all coefficients $\tilde g_{j_1,\dots j_N}$ vanish, because any set of $N$ vectors $\vec w_{k_1}, \dots \vec w_{k_N}$ is then linearly dependent. We can thus assume that $W^\prime$ has full rank, and that the $N$ vectors with odd index, $\vec w_1, \vec w_{2+1}, \dots \vec w_{2(N-1)+1}$, are linearly independent. This assumption does not restrict generality, since simple re-shuffling of the matrix and re-labeling of degrees of freedom can always provide us with such a situation. Consequently, the remaining vectors with even indices ($\vec w_2, \vec w_4,  \dots$) depend linearly on those with odd index, and we can always find coefficients $c_{2k}^{({2(j-1)+1})}$ such that 
\eq 
\forall k  : \vec w_{2k}=\sum_{j=1}^N c_{2k}^{({2(j-1)+1})} \vec w_{2(j-1)+1} \label{linkomm} ,
\en
i.e.~the vectors with even index can be expressed as linear combination of the vectors with odd index. In order to express  (\ref{initialrepredet}) in simpler terms, we use $\text{det}(A B)=\text{det}(A) \text{det}(B)$, or, equivalently, 
\eq 
\text{Det}\left[ \left(\sum_{j=1}^N b_j^{(1)} \vec v_j, \sum_{j=1}^N b_j^{(2)} \vec v_j, \dots \sum_{j=1}^N b_j^{(N)} \vec v_j  \right)  \right] \nonumber \\ = \text{Det}\left[ \left(\begin{array}{cccc} 
b_1^{(1)} & b_2^{(1)} & \dots & b_N^{(1)} \\
\vdots & \vdots & \ddots &\vdots \\
b_1^{(N)} & b_2^{(N)} & \dots & b_N^{(N)} 
\end{array} \right) \right] \nonumber \\  \times \text{Det}\left[\left( \vec v_1 , \vec v_2, \dots \vec v_N \right) \right] . \label{matrixproduct}
\en 
In particular, the coefficient $\tilde g_{1,1,\dots 1}$ is given by the determinant of the matrix consisting of the odd vectors  $\vec w_{1}, \vec w_{3}, \vec w_5 \dots \vec w_{2N-1}$ (see (\ref{gexpress}) and (\ref{initialrepredet})):
\eq 
\tilde g_{1,1\dots 1}=\text{Det}\left[ \left(\vec w_1, \vec w_{3}, \dots \vec w_{2N-1} \right) \right] =: \mathcal{D} .
\en
Since the vectors $\vec w_{2k}$ depend linearly on the $\vec w_{2(j-1)+1}$, we can also express \emph{all} other coefficients $\tilde g_{j_1, \dots j_N}$ as a product of $\mathcal{D}$ with determinants of matrices composed of the coefficients $c^{(m)}_{l}$. Therefore, i.e.~exploiting (\ref{matrixproduct}), we have
\eq 
\tilde g_{1,1\dots 1} &=& \mathcal{D}  , \nonumber  \\
\tilde g_{2,1\dots 1} &=& \mathcal{D} \cdot c_{2}^{(1)} =  \mathcal{D} \cdot \text{Det}\left[ \left( c_2^{(1)} \right) \right] \nonumber  ,\\
\vdots \nonumber \\
\tilde g_{1,\dots,1, 2} &=& \mathcal{D} \cdot c_{2N}^{({2N-1})} = \mathcal{D} \cdot  \text{Det}\left[ \left( c_{2N}^{({2N-1})} \right) \right] , \nonumber \\ 
\tilde g_{2,2,1,\dots, 1} &=&  \mathcal{D} \cdot \left(c_{2}^{(1)} c_{4}^{(3)} -c_{2}^{(3)} c_{4}^{(1)} \right) \nonumber \\ & =&  \nonumber \mathcal{D} \cdot \text{Det}\left[ \left( \begin{array}{cc}  c_{2}^{(1)} & c_{2}^{(3)} \\ c_{4}^{(1)} & c_{4}^{(3)} \end{array} \right) \right] , \\
\vdots \nonumber  \\
\tilde g_{1,\dots 1,2,2} &=& \mathcal{D} \cdot \left(c_{2N-2}^{({2N-3})} c_{2N}^{({2N-1})} - c_{2N-2}^{({2N-1})} c_{2N}^{({2N-3})} \right)  \nonumber \\ &=&   \mathcal{D} \cdot  \text{Det}\left[ \left( 
\begin{array}{cc}  
c_{2N-2}^{({2N-3})} & c_{2N-2}^{({2N-1})} \vspace{2mm} \\  c_{2N}^{({2N-3})} &  c_{2N}^{({2N-1})} \end{array} \right) \right] . \label{detss2}
\en
For a general systematic representation of the $\tilde g_{j_1, \dots, j_N}$ as they are given in (\ref{detss2}), we group the coefficients $c_{2k}^{({2(j-1)+1})}$ in a square $N\times N$ matrix $\mathcal{C}$ with entries
\eq 
\mathcal{C}_{k,j}=c_{2k}^{({2(j-1)+1})} ,
\en
such that the state coefficient $\tilde g_{j_1, \dots j_N}$ is proportional to some \emph{principle minor} (or \emph{coaxial sub-determinant}) of $\mathcal{C}$  \cite{Bernstein:2009uq}, i.e.~the determinant of the sub-matrix $\mathcal{\bar C}(j_1, \dots j_N)$ that is constructed as follows: 
 For each $k$ (with $1\le k \le N$), delete the $k$th row and the $k$th column in the matrix $\mathcal{C}$ if $j_k$ is odd. The matrix $\mathcal{\bar C}(j_1, \dots j_N)$ is thus an $m\times m$-submatrix of $\mathcal{C}$, and $m$ is the number of even coefficients $j_k$, such that $0 \le m \le N$. 

In particular, for $\tilde g_{1,1,1\dots 1}$, all $j_k$ are odd, and $\mathcal{\bar C}(j_1, \dots j_N) $ is the empty $0\times 0$ matrix: All rows and columns have been deleted, and $\tilde g_{1,1,\dots,1}=\mathcal{D}$ -- we set the determinant of the empty $0\times 0$-matrix to unity, for convenience. For $\tilde g_{2,2,2\dots 2}$, one needs to compute the determinant of the full matrix $\mathcal{C}=\mathcal{\bar C}(2,2,\dots , 2)$, since all $j_k$ are even. In general, we can rewrite the coefficients $\tilde g_{j_1, \dots j_N}$ in the state expansion as a product of two determinants, just like in (\ref{matrixproduct}):
\eq 
\tilde g_{j_1, \dots j_N} & = & \mathcal{D} \cdot \text{Det}\left[  \mathcal{\bar C}(j_1, \dots j_N)  \right] . \label{produrepredet}
\en
That is to say, the $2^N$ principle minors of $\mathcal{C}$ (including the empty matrix and the full matrix) are the coefficients in the state representation. Since the values of $\vec w_1, \vec w_3, \dots \vec w_{2N-1}$ only affect the determinant $\mathcal{D}$, they contribute to only one free parameter.  Consequently, at this stage, the maximal number of independent parameters is bounded from above by $N^2+1$ ($N^2$ parameters for the matrix $\bar{ \mathcal{C}}$ and one parameter for $\mathcal{D}$). 

Given the functional form of $\tilde g_{j_1, \dots,j_N}$ in terms of the principle minors of a matrix, we can actually find the exact number of independent parameters: The $2^N$ principle minors can be expressed by only $N^2-N+1$ parameters (a result which dates back to 1893 \cite{MacMahon:1893ff,Stouffer:1924lh,Muir:1894fu}). Consequently, we find \eq \text{dim}(\Xi_{\text{F}}) = N^2-N+2 .\en

 The principle minors also allow us to find a minimal explicit parametrization for accessible fermionic states. Since the set of all $N$ ($1\times 1$)-minors, all $(N(N-1)/2)$ $(2\times 2)$-minors and all $(N-1)(N-2)/2$ $(3\times 3)$-minors that include the first row and the first column are sufficient to span the set of all $2^N-1$ minors \cite{Muir:1894fu}, we can also express all accessible states in terms of these minors. 

\subsection{Fermionic qudits}
In the case $d>2$, we have a $(d\cdot N) \times N$ dimensional matrix $W^\prime$ that determines the scattering process. The vectors $\vec w_2, \dots, \vec w_d, \vec w_{d+2}, \dots, \vec w_{2d},\dots,\vec w_{N d}$ can be expressed as linear combination of the vectors $\vec w_1, \vec w_{d+1}, \dots, \vec w_{(N-1)d+1}$, such that, in analogy to (\ref{linkomm}), we find 
\eq 
\forall k\neq 1+d\cdot l : \vec w_k = \sum_{j=1}^N c_k^{({d(j-1)+1})} \vec w_{d(j-1)+1}  ,
\en 
and the $c_l^{(k)}$ fill a matrix of size $(d-1)N \times N$:
\eq 
\mathcal{C} =
\left(\begin{array}{ccccc}
c_2^{(1)} & c_2^{({d+1})} & c_2^{({2d+1})} & \dots & c_2^{({(N-1)d+1} )}\\
c_3^{(1)} & c_3^{({d+1})} & c_3^{({2d+1})} & \dots & c_3^{({(N-1)d+1})} \\
\vdots &  \vdots& \vdots & \ddots & \vdots \\
c_d^{(1)} & c_d^{({d+1})} & c_d^{({2d+1})} & \dots & c_d^{({(N-1)d+1})} \\ 
c_{d+2}^{(1)} & c_{d+2}^{({d+1})} & c_{d+2}^{({2d+1})} & \dots & c_{d+2}^{({(N-1)d+1})} \\ 
\vdots &  \vdots& \vdots & \ddots & \vdots \\
c_{dN}^{(1)} & c_{d N}^{({d+1})} & c_{d N}^{({2d+1})} & \dots & c_{d N}^{({(N-1)d+1})} \\ 
\end{array} \right) 
\en
In other words, for each additional dimension, we have to add $N$ rows in the coefficient matrix. The $d^N$ coefficients $\tilde g_{j_1,\dots j_N}$ correspond to $d^N$ minors of the matrix $\mathcal{C}$.  Consequently, the accretion of one dimension leads to up to $N^2$ new parameters in the matrix $\mathcal{C}$ and thus to at most $N^2$ new independent parameters. We thus have 
\eq \text{dim}(\Xi_{\text{F}}) \le (d-1)N^2-N+2 ,   \label{dimferapp} \en
which is the general bound for the dimensionality of the manifold of accessible states for $N$ fermions with $d$ internal states, when the fermions start in a non-entangled state.

\subsection{Bosons}
For bosons, the coefficients (\ref{gexpress}) are not the determinant of a sub-matrix of $W^\prime$, but they instead correspond to the \emph{permanent} \cite{Ryser:1963oa,Sachkov:2002fk,Aaronson:2010fk}. Since the permanent of a product of two matrices is not necessarily the product of the permanents of the two matrices, the argument that allowed us to reduce the number of parameters for fermions immediately breaks down for bosons. Consequently, the lack of algebraic symmetry properties of the permanent leads to a larger value for the upper bound to the dimensionality of accessible states for bosons than for fermions. 

In order to see this, we start with the $dN^2$ parameters $W^\prime_{k,l}$ that govern the $d^N$ coefficients $g_{j_1, \dots , j_N}$ of the state. In the following, we identify $2N-1$ parameters that contribute to the normalization and to the global phase of the final state, and thus eliminate $2N-2$ of these. 

Firstly, a complex factor that is applied to each group of $d$ columns leads to a global factor in the final state. More precisely, a factor $\alpha_l$ to the $l$th group of $d$ columns corresponds to 
\eq 
W^\prime_{j, (l-1) \cdot d +k} \rightarrow  \alpha_l  W^\prime_{j,(l-1) \cdot d + k } ,
\en
for $1 \le j \le N$, $1 \le l \le N$, $1 \le k \le d$. By inserting the scaled $W^\prime$ into (\ref{gexpress}), we see that these factors lead to a global factor that only affects the normalization and the phase of the final state, 
\eq 
\tilde g_{j_1, \dots , j_N} \rightarrow \left( \prod_{l=1}^N \alpha_l \right) \tilde g_{j_1, \dots , j_N} .
\en
In close analogy, the application of a factor to each row of $W^\prime$, {\it i.e}~
\eq
W^\prime_{j, (l-1) \cdot d +k} \rightarrow  \beta_j  W^\prime_{j,(l-1) \cdot d + k } ,
\en
also gives a global factor 
\eq 
\tilde g_{j_1, \dots , j_N} \rightarrow \left( \prod_{l=1}^N \beta_l  \right) \tilde g_{j_1, \dots , j_N} .
\en
One of the $N$ factors $\beta_j$ can be absorbed by the $\alpha_k$, such that the $\alpha_j$ and $\beta_k$ only provide $2N-1$ independent parameters. In contrast to the determinant, there are no further symmetry properties that allow us to eliminate other parameters, which leads to the following bound for bosons:
\eq \text{dim}(\Xi_{\text{B}}) \le d N^2 -2N+2 , \en
which is always equal or higher than the corresponding bound for fermions, Eq.~(\ref{dimferapp}).

\FloatBarrier


\begin{thebibliography}{10}

\bibitem{Pan:2011fk}
J.-W. Pan, Z.-B. Chen, C.-Y. Lu, H.~Weinfurter, A.~Zeilinger, and M.~Zukowski,
\newblock Rev. Mod. Phys. {\bf 84}, 777 (2012).

\bibitem{Bell:1964pt}
J.~Bell,
\newblock Physics {\bf 1}, 195 (1964).

\bibitem{Aspect:1982ly}
A.~Aspect, P.~Grangier, and G.~Roger,
\newblock Phys. Rev. Lett. {\bf 49}, 91 (1982).

\bibitem{Aspect:1981zr}
A.~Aspect, P.~Grangier, and G.~Roger,
\newblock Phys. Rev. Lett. {\bf 47}, 460 (1981).

\bibitem{Ekert:1991kx}
A.~K. Ekert,
\newblock Phys. Rev. Lett. {\bf 67}, 661 (1991).

\bibitem{Walther:2005qf}
P.~Walther, K.~J. Resch, T.~Rudolph, E.~Schenck, H.~Weinfurter, V.~Vedral,
  M.~Aspelmeyer, and A.~Zeilinger,
\newblock Nature {\bf 434}, 169 (2005).

\bibitem{PhysRevLett.61.2921}
Y.~H. Shih and C.~O. Alley,
\newblock Phys. Rev. Lett. {\bf 61}, 2921 (1988).

\bibitem{Bouwmeester:1999ys}
D.~Bouwmeester, J.-W. Pan, M.~Daniell, H.~Weinfurter, and A.~Zeilinger,
\newblock Phys. Rev. Lett. {\bf 82}, 1345 (1999).

\bibitem{Eibl:2003ly}
M.~Eibl, S.~Gaertner, M.~Bourennane, C.~Kurtsiefer, M.~\.Zukowski, and
  H.~Weinfurter,
\newblock Phys. Rev. Lett. {\bf 90}, 200403 (2003).

\bibitem{Zhao:2004dz}
Z.~Zhao, Y.-A. Chen, A.-N. Zhang, T.~Yang, H.~J. Briegel, and J.-W. Pan,
\newblock Nature {\bf 430}, 54 (2004).

\bibitem{Lu:2007ve}
C.-Y. Lu, X.-Q. Zhou, O.~Guhne, W.-B. Gao, J.~Zhang, Z.-S. Yuan, A.~Goebel,
  T.~Yang, and J.-W. Pan,
\newblock Nat. Phys. {\bf 3}, 91 (2007).

\bibitem{Radmark:2009ij}
M.~R{\aa}dmark, M.~\.Zukowski, and M.~Bourennane,
\newblock Phys. Rev. Lett. {\bf 103}, 150501 (2009).

\bibitem{Prevedel:2009ec}
R.~Prevedel, G.~Cronenberg, M.~S. Tame, M.~Paternostro, P.~Walther, M.~S. Kim,
  and A.~Zeilinger,
\newblock Phys. Rev. Lett. {\bf 103}, 020503 (2009).

\bibitem{Wieczorek:2009ff}
W.~Wieczorek, R.~Krischek, N.~Kiesel, P.~Michelberger, G.~T\'oth, and
  H.~Weinfurter,
\newblock Phys. Rev. Lett. {\bf 103}, 020504 (2009).

\bibitem{Yao:2011uq}
X.-C. Yao, T.-X. Wang, P.~Xu, H.~Lu, G.-S. Pan, X.-H. Bao, C.-Z. Peng, C.-Y.
  Lu, Y.-A. Chen, and J.-W. Pan,
\newblock Nat. Photon. {\bf 6}, 225 (2012).

\bibitem{Huang:2011fk}
Y.-F. Huang, B.-H. Liu, L.~Peng, Y.-H. Li, L.~Li, C.-F. Li, and G.-C. Guo,
\newblock Nat. Commun. {\bf 2}, 546 (2011).

\bibitem{RevModPhys.81.865}
R.~Horodecki, P.~Horodecki, M.~Horodecki, and K.~Horodecki,
\newblock Rev. Mod. Phys. {\bf 81}, 865 (2009).

\bibitem{Wieczorek:2009fe}
W.~Wieczorek, N.~Kiesel, C.~Schmid, and H.~Weinfurter,
\newblock Phys. Rev. A {\bf 79}, 022311 (2009).

\bibitem{Blatt:2008cl}
R.~Blatt and D.~Wineland,
\newblock Nature {\bf 453}, 1008 (2008).

\bibitem{Popescu:2007kx}
S.~Popescu,
\newblock Phys. Rev. Lett. {\bf 99}, 130503 (2007).

\bibitem{PhysRevLett.82.2594}
J.~Brendel, N.~Gisin, W.~Tittel, and H.~Zbinden,
\newblock Phys. Rev. Lett. {\bf 82}, 2594 (1999).

\bibitem{Marcikic:2004qr}
I.~Marcikic, H.~de~Riedmatten, W.~Tittel, H.~Zbinden, M.~Legr\'e, and N.~Gisin,
\newblock Phys. Rev. Lett. {\bf 93}, 180502 (2004).

\bibitem{Halder:2007th}
M.~Halder, A.~Beveratos, N.~Gisin, V.~Scarani, C.~Simon, and H.~Zbinden,
\newblock Nat. Phys. {\bf 3}, 692 (2007).

\bibitem{PhysRevLett.94.220501}
M.~N. O'Sullivan-Hale, I.~A.~Khan, R.~W. Boyd, and J.~C. Howell,
\newblock Phys. Rev. Lett. {\bf 94}, 220501 (2005).

\bibitem{Vaziri:2002nx}
A.~Vaziri, G.~Weihs, and A.~Zeilinger,
\newblock Phys. Rev. Lett. {\bf 89}, 240401 (2002).

\bibitem{Inoue:2009dq}
R.~Inoue, T.~Yonehara, Y.~Miyamoto, M.~Koashi, and M.~Kozuma,
\newblock Phys. Rev. Lett. {\bf 103}, 110503 (2009).

\bibitem{Mair:2001zr}
A.~Mair, A.~Vaziri, G.~Weihs, and A.~Zeilinger,
\newblock Nature {\bf 412}, 313 (2001).

\bibitem{PhysRevA.83.012306}
L.~Chen and W.~She,
\newblock Phys. Rev. A {\bf 83}, 012306 (2011).

\bibitem{Thew:2004ly}
R.~T. Thew, A.~Ac\'{\i}n, H.~Zbinden, and N.~Gisin,
\newblock Phys. Rev. Lett. {\bf 93}, 010503 (2004).

\bibitem{PhysRevA.69.050304}
H.~de~Riedmatten, I.~Marcikic, V.~Scarani, W.~Tittel, H.~Zbinden, and N.~Gisin,
\newblock Phys. Rev. A {\bf 69}, 050304 (2004).

\bibitem{Wieczorek:2008uq}
W.~Wieczorek, C.~Schmid, N.~Kiesel, R.~Pohlner, O.~G\"uhne, and H.~Weinfurter,
\newblock Phys. Rev. Lett. {\bf 101}, 010503 (2008).

\bibitem{PhysRevA.80.022308}
H.~Hossein-Nejad, R.~Stock, and D.~F.~V. James,
\newblock Phys. Rev. A {\bf 80}, 022308 (2009).

\bibitem{PhysRevLett.99.193602}
C.~Thiel, J.~von Zanthier, T.~Bastin, E.~Solano, and G.~S. Agarwal,
\newblock Phys. Rev. Lett. {\bf 99}, 193602 (2007).

\bibitem{Maser:2009pi}
A.~Maser, U.~Schilling, T.~Bastin, E.~Solano, C.~Thiel, and J.~von Zanthier,
\newblock Phys. Rev. A {\bf 79}, 033833 (2009).

\bibitem{Schilling:2009fk}
U.~Schilling, C.~Thiel, E.~Solano, T.~Bastin, and J.~von Zanthier,
\newblock Phys. Rev. A {\bf 80}, 022312 (2009).

\bibitem{PhysRevLett.102.053601}
T.~Bastin, C.~Thiel, J.~von Zanthier, L.~Lamata, E.~Solano, and G.~S. Agarwal,
\newblock Phys. Rev. Lett. {\bf 102}, 053601 (2009).

\bibitem{Maser:2010fu}
A.~Maser, R.~Wiegner, U.~Schilling, C.~Thiel, and J.~von Zanthier,
\newblock Phys. Rev. A {\bf 81}, 053842 (2010).

\bibitem{Lim:2005bf}
Y.~L. Lim and A.~Beige,
\newblock Phys. Rev. A {\bf 71}, 062311 (2005).

\bibitem{Lutkenhaus:1999bh}
N.~L\"utkenhaus, J.~Calsamiglia, and K.~A. Suominen,
\newblock Phys. Rev. A {\bf 59}, 3295 (1999).

\bibitem{Vaidman:1999qf}
L.~Vaidman and N.~Yoran,
\newblock Phys. Rev. A {\bf 59}, 116 (1999).

\bibitem{Calsamiglia:2002dq}
J.~Calsamiglia,
\newblock Phys. Rev. A {\bf 65}, 030301 (2002).

\bibitem{Aaronson:2010fk}
S.~Aaronson and A.~Arkhipov,
\newblock The computational complexity of linear optics,
\newblock in {\em Proc. 43rd ACM Symp. Theo. Comp.}, STOC '11, p. 333, New
  York, NY, USA, 2011, ACM.


\bibitem{Pittman}
T.B.~Pittman, B.C.~Jacobs, and J.D.~Franson,
\newblock Phys. Rev. A {\bf 64}, 062311 (2001). 

\bibitem{PhysRevA.83.022328}
M.~Huber, H.~Schimpf, A.~Gabriel, C.~Spengler, D.~Bru\ss{}, and B.~C. Hiesmayr,
\newblock Phys. Rev. A {\bf 83}, 022328 (2011).

\bibitem{Bastin:2009ye}
T.~Bastin, S.~Krins, P.~Mathonet, M.~Godefroid, L.~Lamata, and E.~Solano,
\newblock Phys. Rev. Lett. {\bf 103}, 070503 (2009).

\bibitem{PhysRevLett.104.020504}
B.~Kraus,
\newblock Phys. Rev. Lett. {\bf 104}, 020504 (2010).

\bibitem{Bin-Liu1:2011fk}
B.~Liu, J.-L. Li, X.~Li, and C.-F. Qiao,
\newblock Phys. Rev. Lett. {\bf 108}, 050501 (2012).

\bibitem{salwey}
F.~Mintert, B.~Salwey, and A.~Buchleitner,
\newblock Phys. Rev. A {\bf 86},  052330 (2012). 

\bibitem{PhysRevLett.106.180502}
P.~Ribeiro and R.~Mosseri,
\newblock Phys. Rev. Lett. {\bf 106}, 180502 (2011).

\bibitem{1367-2630-14-9-093015}
M.~C. Tichy, M.~Tiersch, F.~Mintert, and A.~Buchleitner,
\newblock N. J. Phys. {\bf 14}, 093015 (2012).

\bibitem{tichydiss}
M.~C. Tichy, {\it Entanglement and interference of identical particles}, PhD thesis, University of Freiburg (2011), www.freidok.uni-freiburg.de/volltexte/8233/

\bibitem{Eisert:2001uq}
J.~Eisert and H.~J. Briegel,
\newblock Phys. Rev. A {\bf 64}, 022306 (2001).

\bibitem{Halevy:2011cr}
A.~Halevy, E.~Megidish, T.~Shacham, L.~Dovrat, and H.~S. Eisenberg,
\newblock Phys. Rev. Lett. {\bf 106}, 130502 (2011).

\bibitem{Monroe}
    S. Olmschenk, D.N. Matsukevich, P. Maunz, D. Hayes, L.-M. Duan, and C. Monroe,
    \newblock Science {\bf 323},  486 (2009) 

\bibitem{Zou:2005hc}
X.~B.~Zou and W.~Mathis,
\newblock Phys. Rev. A {\bf 71}, 042324 (2005).

\bibitem{FRANKEARNOLD208}
S.~Franke-Arnold, L.~Allen, and M.~Padgett,
\newblock Laser \& Photon. Rev. {\bf 2}, 299 (2008).

\bibitem{Tichy:2011fk}
M.~C. Tichy, F.~Mintert, and A.~Buchleitner,
\newblock J. Phys. B {\bf 44}, 192001 (2011).

\bibitem{Ghne:2009ys}
O.~G\"uhne and G.~T\'oth,
\newblock Phys. Rep. {\bf 474}, 1 (2009).

\bibitem{Ghirardi:2003uq}
G.~Ghirardi and L.~Marinatto,
\newblock Fortschr. Phys. {\bf 51}, 379 (2003).

\bibitem{Ghirardi:2004fk}
G.~C.~Ghirardi and L.~Marinatto,
\newblock Phys. Rev. A {\bf 70}, 012109 (2004).

\bibitem{PROP:PROP201200079}
M.~C. Tichy, F.~de~Melo, M.~Ku{\'s}, F.~Mintert, and A.~Buchleitner,
\newblock Fortschr. Phys. {\bf 61}, 225  (2012).

\bibitem{Nielsen:2000fk}
M.~A. Nielsen and I.~L. Chuang,
\newblock {\em Quantum computation and quantum information} (Cambridge
  University Press, Cambridge, 2000).

\bibitem{Werner:1989ve}
R.~F. Werner,
\newblock Phys. Rev. A {\bf 40}, 4277 (1989).

\bibitem{He:2007tg}
B.~He, J.~A. Bergou, and Z.~Wang,
\newblock Phys. Rev. A {\bf 76}, 042326 (2007).

\bibitem{Bernstein:2009uq}
D.~S. Bernstein,
\newblock {\em Matrix Mathematics} (Princeton University Press, Princeton,
  2009).

\bibitem{Ma:2012kx}
X.-S. Ma, S.~Kropatschek, W.~Naylor, T.~Scheidl, J.~Kofler, T.~Herbst,
  A.~Zeilinger, and R.~Ursin,
\newblock Opt. Express {\bf 20}, 23126 (2012).

\bibitem{Nielsen:1999uq}
M.~A. Nielsen,
\newblock Phys. Rev. Lett. {\bf 83}, 436 (1999).

\bibitem{Vidal:2000kx}
G.~Vidal,
\newblock J. Mod. Optic {\bf 47}, 355 (2000).

\bibitem{PhysRevA.83.042332}
D.~J.~H. Markham,
\newblock Phys. Rev. A {\bf 83}, 042332 (2011).

\bibitem{Majorana:1932qa}
E.~Majorana,
\newblock Nuovo Cimento {\bf 9}, 43 (1932).

\bibitem{Sasaki:2011uq}
T.~Sasaki, T.~Ichikawa, and I.~Tsutsui,
\newblock Phys. Rev. A {\bf 83}, 012113 (2011).

\bibitem{Ryser:1963oa}
H.~J. Ryser,
\newblock {\em Combinatorial Mathematics}~The Carus mathematical monographs
  series, Vol. 14 (The Mathematical Association of America, Washington, D.C., 1963).

\bibitem{Sachkov:2002fk}
V.~N. Sachkov and V.~E. Tarakanov,
\newblock {\em Combinatorics of Non-Negative Matrices}, in Transl. of Math. Monogr. 213 (American Mathematical
  Society, Providence, Rhode Island, 2002).


\bibitem{Wieczorek:2009fk}
W.~Wieczorek,
\newblock {\em {Multi-Photon Entanglement -- Experimental Observation,
  Characterization, and Application of up to Six-Photon Entangled States}},
\newblock PhD thesis, Ludwig-Maximilians-Universit\"at M\"unchen, 2009.

\bibitem{Benson:2000db}
O.~Benson, C.~Santori, M.~Pelton, and Y.~Yamamoto,
\newblock Phys. Rev. Lett. {\bf 84}, 2513 (2000).

\bibitem{Santori:2002qf}
C.~Santori, D.~Fattal, J.~Vuckovic, G.~S. Solomon, and Y.~Yamamoto,
\newblock Nature {\bf 419}, 594 (2002).

\bibitem{silberhornneu}
M.~F\"ortsch, J.~F\"urst, C.~Wittmann, D.~Strekalov, A.~Aiello, M.~V. Chekhova,
  C.~Silberhorn, G.~Leuchs, and C.~Marquard,
\newblock A versatile source of single photons for quantum information
  processing,
\newblock arxiv:1204.3056, 2012.

\bibitem{TichyLim}
M.C. Tichy, H.-T. Lim, Y.-S. Ra, F.~Mintert, Y.-H. Kim, and A.~Buchleitner,
\newblock Phys. Rev. A {\bf 83}, 062111 (2011).

\bibitem{Lee:2002fk}
H.~Lee, P.~Kok, N.~J. Cerf, and J.~P. Dowling,
\newblock Phys. Rev. A {\bf 65}, 030101 (2002).

\bibitem{Pryde:2003fk}
G.~J. Pryde and A.~G. White,
\newblock Phys. Rev. A {\bf 68}, 052315 (2003).

\bibitem{Brougham:2011vn}
T.~Brougham, V.~Ko{\v s}t{\'a}k, I.~Jex, E.~Andersson, and T.~Kiss,
\newblock Eur. Phys. J. D {\bf 61}, 231 (2011).

\bibitem{Baek:2007mi}
S.-Y. Baek and Y.-H. Kim,
\newblock Phys. Rev. A {\bf 75}, 034309 (2007).

\bibitem{PhysRevA.78.042321}
S.-Y. Baek, S.~S. Straupe, A.~P. Shurupov, S.~P. Kulik, and Y.-H. Kim,
\newblock Phys. Rev. A {\bf 78}, 042321 (2008).

\bibitem{Lanyon:2008bh}
B.~P. Lanyon, T.~J. Weinhold, N.~K. Langford, J.~L. O'Brien, K.~J. Resch,
  A.~Gilchrist, and A.~G. White,
\newblock Phys. Rev. Lett. {\bf 100}, 060504 (2008).


\bibitem{Ikuta:2011ys} 
R.~Ikuta, T.~Tashima, T.~Yamamoto, M.~Koashi, and N. Imoto, 
\newblock Phys. Rev. A {\bf 83}, 012314 (2011).


\bibitem{Megidish:2012uq}
E.~Megidish, T.~Shacham, A.~Halevy, L.~Dovrat, and H.~S. Eisenberg,
\newblock Phys. Rev. Lett. {\bf 109}, 080504 (2012).

\bibitem{Peyronel:2012fk}
T.~Peyronel, O.~Firstenberg, Q.-Y. Liang, S.~Hofferberth, A.~V. Gorshkov,
  T.~Pohl, M.~D. Lukin, and V.~Vuletic,
\newblock Nature {\bf 488}, 57 (2012).

\bibitem{PhysRevA.77.012104}
M.~Hayashi, D.~Markham, M.~Murao, M.~Owari, and S.~Virmani,
\newblock Phys. Rev. A {\bf 77}, 012104 (2008).

\bibitem{Duerr:2000pj}
W.~D{\"u}r, G.~Vidal, and J.~I. Cirac,
\newblock Phys. Rev. A {\bf 62}, 062314 (2000).

\bibitem{zanthierunpubl}
C.~Ammon, A.~Maser, U.~Schilling, T.~Bastin, and J.~von Zanthier,
\newblock Phys. Rev. A {\bf 86},  052308 (2012).

\bibitem{1367-2630-12-7-073025}
M.~Aulbach, D.~Markham, and M.~Murao,
\newblock N. J. Phys. {\bf 12}, 073025 (2010).

\bibitem{Mathonet:2010fu}
P.~Mathonet, S.~Krins, M.~Godefroid, L.~Lamata, E.~Solano, and T.~Bastin,
\newblock Phys. Rev. A {\bf 81}, 052315 (2010).

\bibitem{Lu:2009zr}
C.-Y. Lu, W.-B. Gao, O.~G\"uhne, X.-Q. Zhou, Z.-B. Chen, and J.-W. Pan,
\newblock Phys. Rev. Lett. {\bf 102}, 030502 (2009).

\bibitem{Lu:2009fk}
C.-Y. Lu, T.~Yang, and J.-W. Pan,
\newblock Phys. Rev. Lett. {\bf 103}, 020501 (2009).

\bibitem{Pan:1998fk}
J.-W. Pan and A.~Zeilinger,
\newblock Phys. Rev. A {\bf 57}, 2208 (1998).

\bibitem{PhysRevA.65.052112}
F.~Verstraete, J.~Dehaene, B.~De~Moor, and H.~Verschelde,
\newblock Phys. Rev. A {\bf 65}, 052112 (2002).

\bibitem{MacMahon:1893ff}
P.~A. MacMahon,
\newblock Philos. Trans. R. Soc. Lond. A {\bf 185}, 111 (1894).

\bibitem{Stouffer:1924lh}
E.~B. Stouffer,
\newblock Trans. Am. Math. Soc. {\bf 26}, 356 (1924).

\bibitem{Muir:1894fu}
T.~Muir,
\newblock Phil. Mag. {\bf 38}, 537 (1894).

\end{thebibliography}
\end{document}